\documentclass[pra,nofootinbib,preprintnumbers,tightenlines,twocolumn,superscriptaddress]{revtex4-1}
\usepackage{braket}
\usepackage{array}
\usepackage{dsfont}
\usepackage{amsmath}
\usepackage{pifont}
\usepackage[normalem]{ulem}
\usepackage{amssymb}
\usepackage{xcolor}
\usepackage{amsfonts}
\usepackage{mathtools}
\usepackage{graphicx}
\usepackage{dcolumn}
\usepackage{bm}
\usepackage{multirow}
\usepackage{hhline}

\usepackage{leftidx}
\usepackage{color}
\usepackage{mathtools}
\usepackage{MnSymbol}
\usepackage[mathscr]{eucal}
\usepackage[german,english]{babel}
\usepackage[capitalize]{cleveref}

\usepackage[utf8]{inputenc}
\usepackage{natbib}
\usepackage{comment}
\newcommand{\abs}[1]{\left| #1 \right|}
\begin{document}

\title{Dynamical excitation processes and correlations of three-body\\ two-dimensional mixtures}

\author{G. Bougas}
\email{gbougas@physnet.uni-hamburg.de}
\affiliation{Center for Optical Quantum Technologies, Department of Physics, University of Hamburg, Luruper Chaussee 149, 22761 Hamburg Germany }

\author{S. I. Mistakidis}
\affiliation{ ITAMP, Center for Astrophysics $\vert$ Harvard $\&$ Smithsonian, Cambridge, MA 02138 USA}
\affiliation{Department of Physics, Harvard University, Cambridge, Massachusetts 02138, USA}

\author{P. Giannakeas}
\affiliation{Max-Planck-Institut f\"ur Physik komplexer Systeme, N\"othnitzer Str.\ 38, D-01187 Dresden, Germany }

\author{P. Schmelcher}
\affiliation{Center for Optical Quantum Technologies, Department of Physics, University of Hamburg, Luruper Chaussee 149, 22761 Hamburg Germany }
\affiliation{The Hamburg Centre for Ultrafast Imaging, University of Hamburg, Luruper Chaussee 149, 22761 Hamburg, Germany}

\begin{abstract}
 A scheme is proposed to dynamically excite distinct eigenstate superpositions in three-body Bose-Fermi mixtures confined in a two-dimensional harmonic trap.
 The system is initialized in a non-interacting state with a variable spatial extent, and the scattering lengths are subsequently quenched spaning the regime from weak to strong interactions. For spatial widths smaller than the three-body harmonic oscillator length,  a superposition of trimers and atom-dimers is dynamically attained, otherwise trap states are predominantly populated, as inferred from the frequency spectrum of the fidelity. 
 Accordingly, the Tan contacts evince the build-up of short range two- and three-body correlations in the course of the evolution. A larger spatial extent of the initial state leads to a reduction of few-body correlations, endowed however with characteristic peaks at the positions of the avoided-crossings in the energy spectra, thereby signalling the participation of atom-dimers. 
 Our results expose ways to dynamically excite selectively trimers, atom-dimers and trapped few-body states characterized by substantial correlations and are likely to be accessible within current experiments.    
\end{abstract}

\maketitle

\section{Introduction}
The appealing feature of ultracold physics is the controllability of interactions that permits to study a plethora of phenomena, such as the formation of droplets~\cite{petrov2015quantum,luo2021new,chomaz2022dipolar} and polarons~\cite{massignan2014polarons,ness2020observation}, and understand in depth the build-up of few- and many-body correlations~\cite{mistakidis2022cold}.
More specifically, the few-body correlations can be quantified by Tan contacts.
These stem from the short-range character of the interatomic interactions \cite{tan_energetics_2008,tan_generalized_2008,tan_large_2008,braaten_universal_2011,werner_general_2012,valiente_universal_2012,olshanii_short-distance_2003,He_Concept_2016,Zhang_contact_2017}, and are experimentally probed through radio-frequency (rf) spectroscopy~\cite{fletcher_two-_2017,sagi_measurement_2012,wild_measurements_2012}, time-of-flight expansion~\cite{stewart_verification_2010} or Bragg spectroscopy~\cite{hoinka_precise_2013,kuhnle_universal_2010}. 
Contacts interrelate macroscopic observables at equilibrium, such as energy and pressure of a gas, in terms of few-body microscopic mechanisms ~\cite{werner_general_2012,braaten_universal_2008} addressing the properties of a gas universally regardless of the atom number, the statistics, or the interaction strength.

The recent realization of three-dimensional (3D) unitary Bose gases offers the possibility to investigate the dynamical formation of few-body correlations in strongly interacting ultracold matter \cite{klauss_observation_2017,makotyn_universal_2014,eigen_universal_2017,eigen_universal_2018}.
Quenching the scattering length from the non-interacting case to unitarity enables the experimental observation of few-body states such as the Efimov ones, i.e. an infinite geometric progression of three-body bound levels comprised of unbound two-body subsystems \cite{nielsen_three-body_2001,Efimov_hard-core_1970}.
In addition, theoretical efforts demonstrated that the quenched dynamics of such three-body systems exhibits unique features in the population growth of Efimov trimers and atom-dimers \cite{colussi_dynamics_2018,dincao_efimov_2018,colussi_bunching_2019,colussi_cumulant_2020,Musolino_Bose_2022}.
For example, in Ref.~\cite{colussi_dynamics_2018} it was argued that the two-body Tan contact is enhanced during the early stages of the dynamics, whereas the three-body contact increases appreciably only when the interparticle spacing matches the size of an Efimov state~\cite{dincao_efimov_2018}.
However, the latter are typically short-lived due to three-body recombination processes~\cite{greene_universal_2017}. 

Promising candidates to mitigate such losses maintaining a high fraction of trimer states, are two-dimensional (2D) gases~\cite{dincao_ultracold_2015,levinsen_efimov_2014}. 
There the corresponding trimer wave functions have a small amplitude at short distances suppressing three-body recombination processes~\cite{kirk_three-body_2017,levinsen_efimov_2014} as compared to 3D systems. 
Additionally, theoretical studies in 2D three-body systems \cite{bellotti_mass-imbalanced_2013,pricoupenko_universal_2010,bellotti_scaling_2011,Bougas_Few_2021,sandoval_mean-square_2016,kartavtsev_universal_2006,bruch_binding_1979} have addressed their time-independent attributes in terms of their eigenspectrum as well as their corresponding few-body correlations via Tan contacts \cite{bellotti_contact_2014,valiente_universal_2011,bellotti_dimensional_2013,Bougas_Few_2021}.
In particular, it was shown that mass-imbalanced mixtures support a multitude of trimer states with amplified two- and three-body correlations compared to the mass-balanced case~\cite{Bougas_Few_2021,bellotti_contact_2014}.

In contrast to the 3D systems \cite{kerin_quench_2022,colussi_bunching_2019,dincao_efimov_2018,colussi_dynamics_2018,Kerin_effects_2022}, the dynamical response including the underlying excitation processes and accompanying correlation mechanisms of 2D three-body systems is not well-understood. 
Importantly, the study of these systems is thus far restricted 
to their stationary correlation properties~\cite{bellotti_contact_2014,bellotti_mass-imbalanced_2013,bellotti_scaling_2011,valiente_universal_2011}  in the absence of external confinement. 
In this work, a protocol is proposed for triggering specific excitation branches in 2D harmonically trapped mixtures of two identical bosons or fermions interacting with another atom. 
Apart from the particle statistics, our study addresses the effect of unequal massed three-body collisions. 
In 3D gases, it is known that highly mass-imbalanced systems exhibit rich resonant effects \cite{Ulmanis_heteronuclear_2016,Giannakeas_assymetric_2021,giannakeas_ultracold_2018,Mikkelsen_three-body_2015,wacker_universal_2016,Johansen_testing_2017}, or favor the observation of multiple successive Efimov states~\cite{Tung_Geometric_2014,Pires_Observation_2014,Kerin_Quantum_2022}, while offering unique platforms to study reaction rates in atom-dimer and molecule-molecule collisions \cite{gao_atom_2018,schalchi_cold_2020,schalchi_scattering_2018,rui_controlled_2017,yang_observation_2019,hoffmann_reaction_2018,makrides_collisions_2020}.
Therefore, the inclusion of unequal masses here provides a comprehensive description of the dynamic properties of 2D three-body collisions ranging from light-light-heavy (LLH) to heavy-heavy-light (HHL) systems.

Initially, the three-body mixture is considered in a non-interacting state characterized by a parameter $w$, describing its spatial extent.
Subsequently, the interactions are turned on abruptly (interaction quench), resulting into distinct dynamical response regimes characterized by specific excitation mechanisms and correlations being imprinted in the fidelity spectrum.
The Hilbert space of the post-quench three-body system, at the final values of the scattering lengths, is mainly partitioned into three generic types of eigenstates: trimers, atom-dimers and trap states.
For widths $w$ of the initial state smaller than the harmonic oscillator length scale,
we observe that the dominant excitation branches identified in the fidelity spectrum correspond to trimers and atom-dimers. 
In the case of HHL systems, however, these states are prevalent over a relatively smaller range of scattering lengths.
For an increasing width $w$ of the initial state the trap states are predominantly populated.

Furthermore, we show that the participation of distinct eigenstates impacts strongly the dynamics of short-range correlations quantified by the Tan contacts. 
In particular, both the two- and three-body correlations become enhanced for initial state widths smaller than the spatial extent of the trap.
The correlations are suppressed as the width of the initial configuration is increased since the population of trap states becomes more dominant.
In addition, distinct peaks in the few-body correlations are observed as the scattering lengths vary. 
This structure arises from the narrow avoided-crossings in the eigenspectrum where the atoms are in a superposition of trap and atom-dimer states. 
The above mentioned features occur for both LLH and HHL settings regardless the exchange symmetry of the particles.
However, the enhancements in the few-body contacts become narrower in the HHL case, as compared to the LLH one, due to the existence of sharp avoided-crossings in the respective energy spectrum~\cite{Bougas_Few_2021}. 

This work proceeds as follows: In Sec.~\ref{Sec:Hyper} the adiabatic hyperspherical formalism is briefly outlined, and in Sec.~\ref{Sec:Setup} the initial ansatz of the three-body system and the time-evolved wave function are introduced. Subsequently, the excitation spectra, associated modes and correlation dynamics based on the fidelity spectrum and Tan contacts are unveiled for both LLH systems in Sec.~\ref{Sec:Dyn_LLH} and HHL ones in Sec.~\ref{Sec:Dyn_HHL}. 
In Sec.~\ref{Sec:Exp} we briefly comment on the possible experimental realization of our setup. 
Sec.~\ref{Sec:Outlook} lays out our conclusions and provides an outlook. 
Moreover, Appendix~\ref{Ap:Interactions} introduces the adiabatic Hamiltonian and the 2D zero-range pseudopotential. 
Appendix~\ref{Ap:Hyper_wave} provides the form of the hyperangular wave function for the non-interacting initial state. 
In Appendix~\ref{Ap:Same_width}, we elaborate on the excitation spectrum of the BBX LLH system for widths of the initial state equal to the three-body harmonic oscillator length.

\section{Adiabatic hyperspherical representation of the three-body mixture} \label{Sec:Hyper}

In the following we consider three-body binary mass-imbalanced mixtures trapped in a 2D harmonic oscillator of frequency $\omega$.
They typically consist of either two identical bosons (BBX) or two identical non-interacting spin polarized fermions (FFX) interacting with a third distinguishable particle. 
The underlying pairwise interactions are modeled with $s$-wave zero-range pseudopotentials~\cite{olshanii_rigorous_2001} characterized by 2D scattering lengths $a_{FX}$ and $a_{BB}$, $a_{BX}$ for the FFX and BBX systems, respectively. Here, $a_{\sigma \sigma'}$ denotes the 2D scattering length between a particle of species $\sigma$ and $\sigma'$, where $\sigma=$B, X or $\sigma=$F, X.
Below, for simplicity, we typically consider variations of $1/a_{FX}$ and $a_{BB}/a_{BX}$ where in the latter case $a_{BB}$ is kept fixed.  
The magnitude of the 2D scattering lengths can in principle be adjusted via standard Fano-Feshbach resonances \cite{chin_feshbach_2010}, since they parametrically depend on their 3D counterparts \cite{Petrov_interatomic_2001}. Let us note that by definition the 2D scattering lengths can only be positive, a property stemming from the existence of a two-body bound state always in 2D, and the non-interacting limit occurs when they are either $0$ or $+\infty$ \cite{liu_exact_2010}.
Moreover, depending on the mass ratio between the identical atom and the third particle, i.e. $m_{B/F}/m_X$, we  distinguish between 
LLH and HHL cases. In particular, the employed 
mass ratios are $m_B/m_X=0.04, \, 22.16$ for BBX 
referring to mixtures of $^7\rm{Li}-^7\rm{Li}-^{173}\rm{Yb}, \: ^{133}\rm{Cs}-^{133}\rm{Cs}-^6\rm{Li}$ and $m_F/m_X=0.0451,\, 24.71$ for FFX corresponding to $^6\rm{Li}-^6\rm{Li}-^{133}\rm{Cs}, \: ^{173}\rm{Yb}-^{173}\rm{Yb}-^7\rm{Li}$ systems.

The stationary properties of these mixtures are straightforwardly addressed within the adiabatic hyperspherical framework~\cite{nielsen_three-body_2001,greene_universal_2017,naidon_efimov_2017,dincao_few-body_2018,rittenhouse_greens_2010}, with the pairwise interactions modeled via contact pseudopotentials. 
Owing to the decoupling of the center of mass, the hyperspherical coordinates representation is employed and the relative position of the atoms is described by a set of three hyperangles (which collectively are denoted by $\boldsymbol{\Omega}$) and the hyperradius $R$ that controls the overall size of the system.
Hence, by employing the hyperspherical coordinates the relative three-body Hamiltonian~\cite{Bougas_Few_2021} reads:
\begin{equation}
	     H_{\rm{rel}}=-\frac{\hbar^2}{2\mu R^{3/2}}\frac{\partial^2}{\partial R^2} R^{3/2}+ \frac{1}{2}\mu \omega^2 R^2+ H_{{\rm ad}}(R;\boldsymbol{\Omega}).
	     \label{Eq:hamilt_Hyper}
\end{equation}
The first term refers to the kinetic energy, while the second one is the external trapping potential. $H_{{\rm ad}}(R;\boldsymbol{\Omega})$ describes the centrifugal motion of the three particles, and contains the pairwise $s$-wave contact interactions, depending on the aforementioned 2D scattering lengths [for more details see Appendix \ref{Ap:Interactions}].
Also, $\mu=m_{B/F}/\sqrt{2m_{B/F}/m_X+1}$ is the three-body reduced mass and $m_{B/F}$ stands for the mass of bosons or fermions. 
Note that in the following we employ as a characteristic length scale of the three-body system the quantity $a_{\rm{ho}}=\sqrt{\hbar/\mu \omega}$, i.e. {\it the three-body harmonic oscillator length}.

The eigenstates of the three-body system are determined as follows: First, $H_{{\rm ad}}(R;\boldsymbol{\Omega})$ is diagonalized at fixed hyperradius $R$~\cite{rittenhouse_greens_2010} where the eigenvalues $s_\nu(R)$ are associated with the adiabatic potential curves $\hbar^2(s_{\nu}^2(R)-1/4)/2\mu R^2$ and the corresponding eigenfunctions, i.e. $\Phi_{\nu}(R;\boldsymbol{\Omega})$, are used as basis set for the three-body relative wave function.
The latter in the adiabatic hyperspherical representation is given by the expression $\Psi(R,\boldsymbol{\Omega})=R^{-3/2} \sum_{\nu} F_{\nu}(R) \Phi_{\nu}(R;\boldsymbol{\Omega})$ \footnote{We note that in the following sections and appendices the wave functions with the superscripts $\Psi(R,\boldsymbol{\Omega},t)$ or $\Psi^{\rm{f}}(R,\boldsymbol{\Omega})$ indicate the time-evolved wave function at time $t$ or the post-quench $\rm{f}$-th eigenstate respectively [see also Sec. \ref{Sec:Setup}].}.
$F_{\nu}(R)$ denotes the hyperradial component of $\Psi(R,\boldsymbol{\Omega})$ which satisfies  the following system of coupled ordinary differential equations:

\begin{eqnarray}
  & &  \Big\{ -\frac{\hbar^2}{2\mu}\frac{d^2}{d R^2}+U_{\nu}(R)  \Big\} F_{\nu}(R) \nonumber \\
& &    -\frac{\hbar^2}{2\mu} \sum_{\nu'} \left[2P_{\nu \nu'}(R) \frac{d}{dR}+Q_{\nu \nu'}(R) \right] \,F_{\nu'}(R)=EF_{\nu}(R). \nonumber \\
\label{Eq:Hyperradial}
\end{eqnarray}
Here, $U_{\nu}(R)$ represents the $\nu$-th adiabatic potential curve including the trap, whereas the $P_{\nu \nu'}(R)$ and $Q_{\nu \nu'}(R)$ terms denote the non-adiabatic coupling matrix elements.
More specifically, the adiabatic potential curves and the non-adiabatic coupling matrix elements are given by the following expressions~\cite{Bougas_Few_2021,rittenhouse_greens_2010,kartavtsev_universal_2006},
\begin{eqnarray}
  U_{\nu}(R)&=&\frac{\hbar^2}{2\mu R^2}\left(s^2_{\nu}(R)-\frac{1}{4} \right)+\frac{1}{2}\mu \omega^2 R^2 
  \label{Eq:Pots_adiabatic}\\
  P_{\nu \nu'}(R)&=&\braket{\Phi_{\nu}(R;\boldsymbol{\Omega})|\frac{\partial \Phi_{\nu'}(R;\boldsymbol{\Omega})}{\partial R}}_{\boldsymbol{\Omega}}  \label{Eq:P_coup}\\
  Q_{\nu \nu'}(R)&=&\braket{\Phi_{\nu}(R;\boldsymbol{\Omega})|\frac{\partial^2 \Phi_{\nu'}(R;\boldsymbol{\Omega})}{\partial R^2}}_{\boldsymbol{\Omega}}, \label{Eq:Q_coup}
\end{eqnarray}
where the symbol $\braket{\ldots}_{\boldsymbol{\Omega}}$ indicates that the integration is over the hyperangles only.
In the following, harmonic oscillator units are adopted, unless stated otherwise, i.e. $m_{B/F}=\hbar=\omega=1$, where $m_{B/F}$ is the mass of the identical bosons or spin polarized fermions.

\section{Initialization and quench protocol}  \label{Sec:Setup}

Initially the three atoms are prepared in a non-interacting state. This situation in 2D translates to a scattering length either $0$ or $+\infty$, which in the case of two harmonically trapped atoms is shown to reproduce the corresponding non-interacting energy spectra~\cite{liu_exact_2010,Bougas_analytical_2019,Busch_two_1998}. The state is characterized by $1/a_{BX}=1/a_{BB}=0$ for BBX or $1/a_{FX}=0$ for FFX systems, while its spatial extent is parametrized by $w$, see \cref{Fig:sketch} (a). 
The initial three-body wave function in the hyperspherical coordinate frame reads 
\begin{eqnarray}
    \Psi(R,\boldsymbol{\Omega},t=0)&=&\frac{R^L\sqrt{2}}{\sqrt{\Gamma(2+L)}w^{2+L}} e^{-\frac{R^2}{2w^2}} \Phi^{(0)}_0(\boldsymbol{\Omega}),
    \label{Eq:Non_int_Gauss_Hyper}
\end{eqnarray}
where $\Gamma(\cdot)$ is the gamma function. 
Also, $\Phi^{(0)}_0(\boldsymbol{\Omega})$ is the non-interacting ground state of $H_{{\rm ad}}(R;\boldsymbol{\Omega})$ [Eq.~\eqref{Eq:hamilt_Hyper}] 
(denoted by the $(0)$ superscript)  taking into account the total angular momentum $L$ and parity $\pi$ of the system $L^{\pi}$. In particular, $L^{\pi}=0^+$ [$L^{\pi}=1^-$] for BBX [FFX] systems. The independence of $\Phi^{(0)}_0(\boldsymbol{\Omega})$ on $R$ stems from the independence of the hyperangular eigenvalues of the non-interacting adiabatic Hamiltonian on this parameter, for more details see Appendix~\ref{Ap:Hyper_wave}. 
The hyperradial part of $\Psi(R,\boldsymbol{\Omega},t=0)$ is the ground state of  the hyperradial equation [Eq.~\eqref{Eq:Hyperradial}] with zero non-adiabatic coupling matrix elements, due to the independence of $\Phi_0^{(0)}(\boldsymbol{\Omega})$ on $R$, and one potential curve, associated to this ground hyperangular state, $U(R)=1/(2\mu R^2)[(L+1)^2-1/4]+1/(2\mu w^4)R^2$. 
Its energy reads $(2+L)/(\mu w^2)$, where $L=0\, (1)$ refers to the total angular momentum for the BBX (FFX) system.

It should be noted that Eq. \eqref{Eq:Non_int_Gauss_Hyper} is an eigenstate of the non-interacting Hamiltonian \cref{Eq:hamilt_Hyper} only in the case of $w=a_{\rm{ho}}$ coinciding with the non-interacting ground trap state.
The spatial extent $w$ can be adjusted experimentally e.g. by means of a trap frequency quench [see \cref{Sec:Exp} for a more detailed discussion], however in the following we treat it as a free parameter. 
This permits us to investigate the role of the spatial extent of the initial wave function on the post-quench dynamics.
However, a detailed argumentation on the interval of values of the width $w$ is provided in Sec.~\ref{Sec:Exp}.
Nevertheless, for typical LLH settings that we shall consider below these bounds yield, $w \geq 0.46$ while for HHL ones, $w \geq 1.16$.

\subsection{Time-evolution of the wave function}\label{time_evol_wfn} 

\begin{figure}[t!]
\centering
\includegraphics[width=0.48 \textwidth]{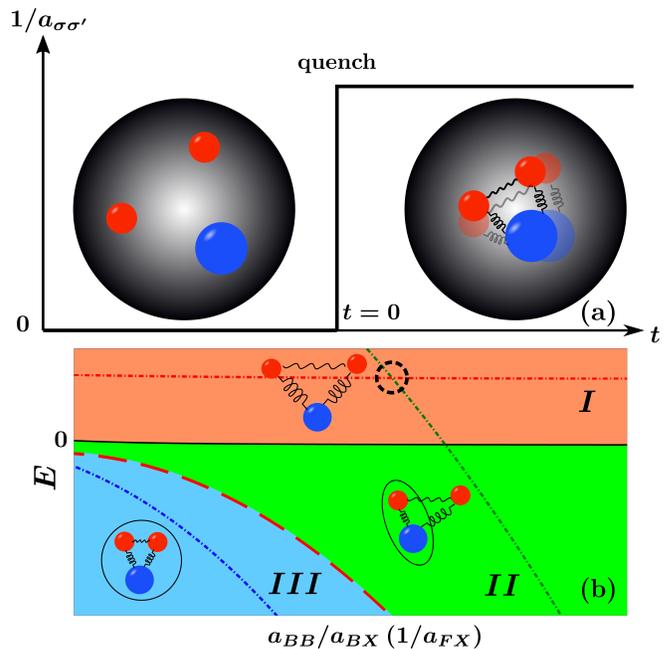}
\caption{(a) Cartoon of the quench scenario. 
The system consists of either two (red particles) identical bosons (BBX) or fermions (FFX) and a distinguishable atom (blue particle). 
They are initialized ($t=0$) in a non-interacting state with spatial extent $w$.
The dynamics is induced by a sudden change of the scattering lengths (interspecies denoted by springs and intraspecies by wiggly lines) from their non-interacting to finite values.
(b) Schematic representation of a typical three-body energy spectrum. In region III, below the BX or FX bare dimer threshold, (red dashed line), trimer states can be formed, denoted by a circle. Region II signals the presence of atom-dimers (dimers are marked by an ellipse), and in region I, trap states appear along with atom-dimers. 
These two latter eigenstates feature avoided-crossings, see for instance the dashed circle. The energy dependence of the trimers, atom-dimers and trap states on the scattering length is schematically presented by the blue, green and red dash-dotted lines respectively. 
Note that the horizontal axis corresponds to a wide range of considered scattering lengths, but does not however reach the zero limit.}
    \label{Fig:sketch}
\end{figure}

To trigger the nonequilibrium dynamics of the three-body mixture we perform quenches of the relevant 2D scattering lengths $a_{\sigma,\sigma'}$. 
Accordingly, their values are suddenly reduced at $t=0$ from their initial non-interacting ones. 
Recall that this is experimentally feasible via appropriate Feshbach resonances [for more details see also \cref{Sec:Exp}].
Specifically, a different quench scheme is applied for the BBX and FFX systems since the former (latter) possesses two (one) scattering lengths, i.e. $a_{BB}$ and $a_{BX}$ ($a_{FX}$).
In the case of FFX mixtures, solely $1/a_{FX}$ is quenched and the consequent dynamics is explored over a wide range of post-quench $1/a_{FX}$  [\cref{Fig:sketch} (a)].  
On the other hand, for the BBX system both the $1/a_{BB}$ and $1/a_{BX}$ are changed abruptly at $t=0$ from their non-interacting values [\cref{Fig:sketch} (a)] towards different post-quench $1/a_{BX}$ and fixed $1/a_{BB}=1$. 
It is worth mentioning that by tuning the magnetic field for the quench in the experiment, both $a_{BX}$ and $a_{BB}$ are affected, and hence broad (narrow) intraspecies (interspecies) resonances are required such that the variation of $a_{BB}$ is very small compared to that of $a_{BX}$ [see also \cref{Sec:Exp}]. 
We remark that $a_{BB}=1$ is chosen such that the bosonic atoms have an intermediate repulsive interaction strength \footnote{The two-body interaction strength between the $\sigma=B,F$ and $\sigma'$ species~\cite{Busch_two_1998,Bougas_analytical_2019,Doganov_two_2013} 
is defined as $g_{\sigma \sigma'}=\left[ \ln\left( 2e^{-2\gamma}(1+m_{\sigma}/m_{\sigma'})/a^2_{\sigma \sigma'}  \right) \right]^{-1}$, where $\gamma=0.577$. This implies that when $a_{\sigma \sigma'}> (<) e^{-\gamma}\sqrt{2}\sqrt{1+\frac{m_{\sigma}}{m_{\sigma'}}}$ attractive (repulsive) effective interaction regimes arise.}.
However, we have checked that the dynamical processes and response of the LLH and HHL BBX systems that are presented below [\cref{Sec:Dyn_LLH}, \cref{Sec:Dyn_HHL}] do not change substantially closer to the non-interacting limit, i.e. $a_{BB}>1$. The fact that the qualitative features of the results remain the same towards the non-interacting limit permits us to expose the role of the particle statistics between BBX and FFX systems. 

To describe the quenched dynamics of the three-body system, the time-evolved wave function is expressed as a projection of the initial state [Eq.~\eqref{Eq:Non_int_Gauss_Hyper}] onto the interacting eigenstates of the post-quench 2D scattering lengths. Specifically, it acquires the form 
\begin{equation}
    \Psi(R,\boldsymbol{\Omega},t)=\sum_{\rm{f}} e^{-i E_{\rm{f}} t} c_{\rm{f},\rm{in}} \Psi^{\rm{f}}(R,\boldsymbol{\Omega}),
    \label{Eq:Quench}
\end{equation}
where $\Psi^{\rm{f}}(R,\boldsymbol{\Omega})=R^{-3/2}\sum_{\nu} F^{\rm{f}}_{\nu}(R)\Phi_{\nu}(R;\boldsymbol{\Omega})$ are the post-quench interacting eigenstates and $E_{\rm{f}}$ their eigenenergies. 
Also, $c_{\rm{f},\rm{in}}=\int dR d\boldsymbol{\Omega}\, R^3 \Psi(R,\boldsymbol{\Omega},t=0) \left[\Psi^f(R,\boldsymbol{\Omega})\right]^*$ denote the overlap coefficients between the initial and the post-quench eigenstates.  
The overlap coefficients are explicitly determined by the initial state and hence its width $w$ for a fixed post-quench scattering length. 
This leads to a $w$-dependent participation of specific post-quench eigenstates , i.e. depending on $w$ different eigenstates contribute in the dynamics, whose distinct features dictate the dynamical response of the system, as it will be demonstrated below [\cref{Sec:Dyn_LLH} and \cref{Sec:Dyn_HHL}].

\subsection{Classification of post-quench three-body eigenstates}\label{classification} 

A detailed knowledge of the three-body energy spectra~\cite{gharashi_three_2012,Bougas_Few_2021,liu_exact_2010}, will allow an in-depth understanding of the emergent nonequilibrium dynamics of both the BBX and FFX mixtures. 
The post-quench interacting eigenstates can be categorized into the so-called trimers, atom-dimers and trap states~\cite{blume_three_2002,portegies_efimov_2011}. 
Trimers are three-body bound states which exist below the BX or FX dimer energies, see in particular the red-dashed line and region III in \cref{Fig:sketch} (b). 
In Ref.~\cite{dincao_ultracold_2015} it was shown that in the absence of a trap the BX or FX dimer energy is given by $E_{\sigma X}=-2e^{-2\gamma}(1+\mathcal{M})/a_{\sigma X}^2$. Here, $\sigma=B,F$,  $\gamma=0.577$ and $\mathcal{M}=m_{\sigma}/m_X$. 
For BBX systems there is also the BB dimer energy determined by $E_{BB}=-4e^{-2\gamma}/a_{BB}^2$ which is constant since $a_{BB}=1$ remains fixed for all the post-quench $a_{BX}$ scattering lengths\footnote{These relations are altered in the presence of a trap only for scattering lengths comparable to or larger than the length scale $\ell=\sqrt{\hbar/\mu_{2B}\omega}$ (with $\mu_{2B}$ being the two-body reduced mass)~\cite{Idziaszek_Analytical_2006}. This effect depends also on the mass ratio of the three-body system.}.

Region II of \cref{Fig:sketch} (b) indicates the energies of the atom-dimer states which are two-body bound states interacting with a third particle. 
The atom-dimer states depend strongly on $a_{BB}/a_{BX}$  [$1/a_{FX}$] in the case of the BBX [FFX] systems having a BX+B [FX+F] character. 
Moreover, the region I of \cref{Fig:sketch} (b) depicts the energy regime of the trap states that are almost insensitive to scattering length variations [see straight lines in \cref{Fig:sketch}(b)] referring to three weakly interacting particles.
Apparently, avoided-crossings occur between BX+B or FX+F atom-dimers, also encountered in region I, and trap states, designated by dashed circles in \cref{Fig:sketch} (b). 
For BBX systems, apart from the aforementioned states appearing in region I, BB+X atom-dimers arise as well. 
Their eigenenergies experience only small variations with respect to $a_{BB}/a_{BX}$, similarly to the trap states, since the post-quench $a_{BB}$ is kept fixed.
A way to distinguish them from trap states is by inspecting their stationary two-body BB short-range correlations, e.g. through the two-body BB contact.  
In Ref.~\cite{Bougas_Few_2021} it was shown that the latter is more pronounced in the case of BB+X atom-dimers than for trap states.

Notably, all three types of eigenstates display a different spatial extent in terms of the hyperradius $R$. 
Therefore, the initial state described by Eq.~\eqref{Eq:Non_int_Gauss_Hyper} will eventually screen out particular states or superpositions in the time-evolution for different widths $w$ and this information is encoded in the overlap coefficients $c_{\rm{f},\rm{in}}$ [see also Sections~\ref{Sec:Dyn_LLH} and~\ref{Sec:Dyn_HHL}].

\begin{figure}[t!]
\centering
\includegraphics[width=0.48 \textwidth]{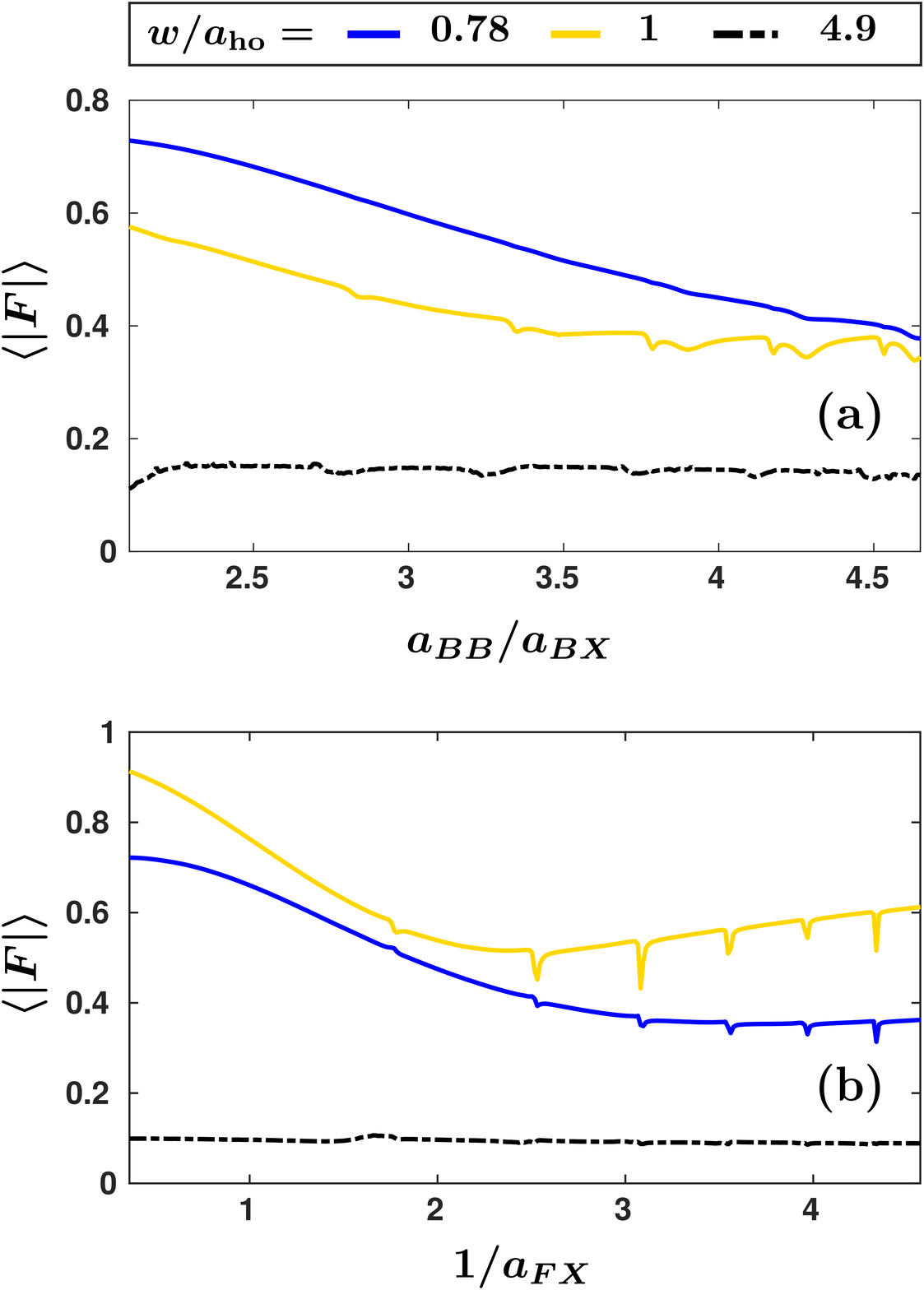}
\caption{Average dynamical response, as captured by the time-averaged fidelity $\langle \abs{F} \rangle$, with respect to the scattering length ratio (a) $a_{BB}/a_{BX}$ for the BBX and (b) $1/a_{FX}$ for the FFX LLH systems. Cases of different widths $w$ (see legend) of the initial state are presented. Apparently, in both settings the response is changed for widths smaller or larger than $a_{\rm{ho}}=1.02$. In particular, it is enhanced for wider initial states having $w>a_{\rm{ho}}$.}
\label{Fig:Fid_Av_LLH}
\end{figure}

\section{Quench dynamics of LLH settings}\label{Sec:Dyn_LLH}

To obtain an overview of the system's dynamical response for different widths of the initial state and post-quench scattering lengths, we employ the time-averaged fidelity~\cite{Bougas_analytical_2019,Bougas_stationary_2020,Budewig_Quench_2019} 
\begin{equation}
\langle |F| \rangle= \lim_{T\to \infty} \frac{\int_0^T dt \,  |F(t)|}{T}.
\label{Eq:Fid_av}
\end{equation}
The total time-evolution $T$ is considered to be long enough such that $\langle |F| \rangle$ is converged.\footnote{Here we consider total evolution times $T=800$, while the time-averaged fidelity for the LLH (HHL) settings saturates already from $T=300$ ($T=500$).}
The fidelity, which essentially estimates the deviation of the time-evolved state [Eq. \eqref{Eq:Quench}] from the initial one, reads  
\begin{eqnarray}
F(t)&=&\langle \Psi(R,\boldsymbol{\Omega},t) |  \Psi(R,\boldsymbol{\Omega},t=0) \rangle \nonumber \\
&=&\sum_{\rm{f}} \abs{c_{\rm{f},\rm{in}}}^2 e^{-i E_{\rm{f}}t}. 
\label{Eq:Fid}
\end{eqnarray}
Here, $c_{\rm{f},\rm{in}}$ are the overlap coefficients introduced in Eq.~(\ref{Eq:Quench}) and $E_{\rm{f}}$ refer to the energies of the post-quench eigenstates. 
As a function of the post-quench scattering length, the dynamical response of the three-body system exhibits two distinct regimes mainly determined by the width of the initial state with respect to the three-body harmonic oscillator length, $a_{\rm{ho}}$. 
In this section the LLH setups that are considered have a mass ratio $m_{B/F}/m_X=0.04$ yielding a three-body harmonic oscillator length $a_{\rm{ho}}=1.02$.

\begin{figure*}[t!]
\centering
\includegraphics[width=1 \textwidth]{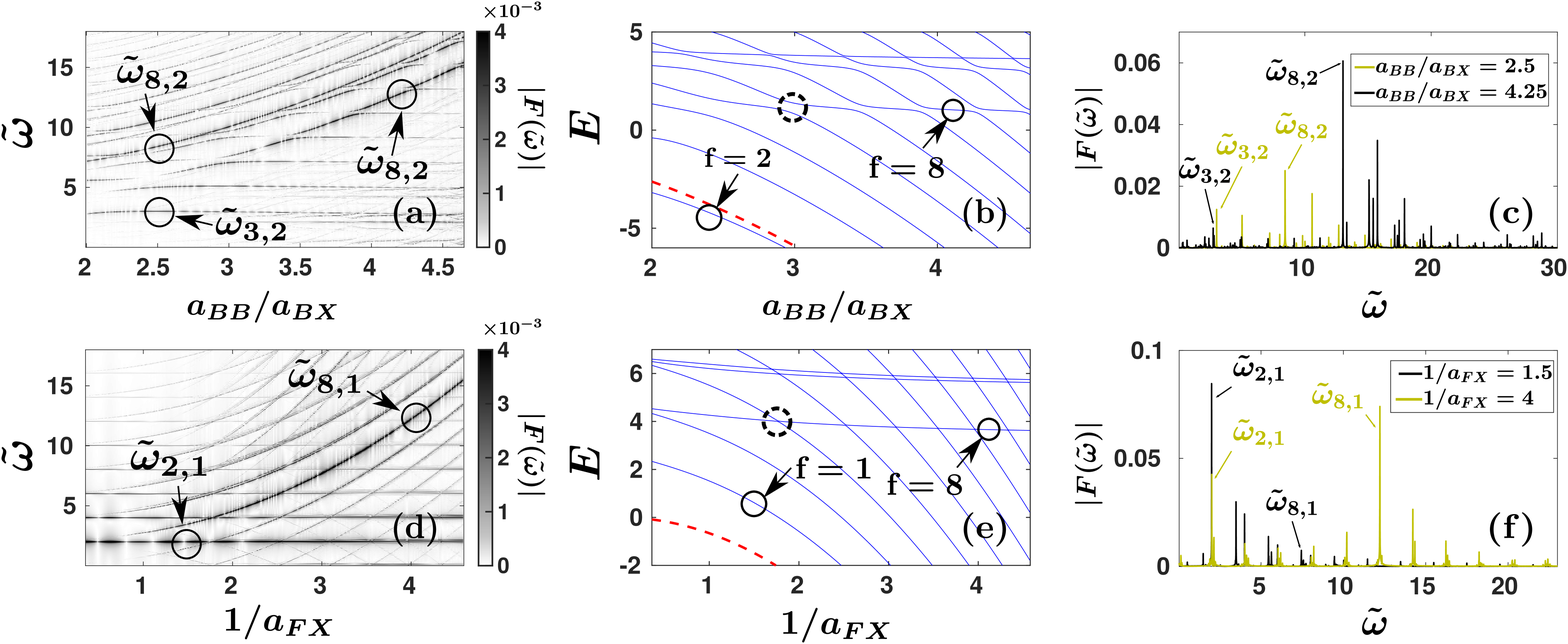}
\caption{Fidelity spectra of the quenched (a) BBX and (d) FFX LLH systems with a narrow pre-quench state of $w/a_{\rm{ho}}=0.78$. The circles denote frequencies associated to specific post-quench eigenstates. 
The interaction-dependent excitation branches signal the dominant participation of trimer and atom-dimer states in the dynamics and refer to their energy differences with respect to trap states. Almost constant branches are related to trap excitations.  
The energy spectra of the (b) BBX and (e) FFX LLH systems, where a series of avoided-crossings among atom-dimers and trap states occurs, marked by the dashed circles. The red dashed line indicates the bare BX or FX dimer threshold. (c), (f) Profiles of the fidelity spectrum for the (c) BBX and (f) FFX mixture at different scattering lengths (see legend).}
\label{Fig:Fid_Spec_LLH}
\end{figure*}

Regarding the LLH BBX system, the time-averaged fidelity $\langle |F| \rangle$ in terms of $a_{BB}/a_{BX}$ is depicted in \cref{Fig:Fid_Av_LLH} (a) for various widths of the initial state. 
Apparently, the qualitative behavior of $\langle |F| \rangle$ depends strongly on $w$. 
For instance, in the case of $w/a_{\rm{ho}}=0.78$ the deviation from the initial state becomes larger for increasing $a_{BB}/a_{BX}$. 
Such a decrease of $\langle |F| \rangle$ holds also when $w/a_{\rm{ho}}=1$ as long as $a_{BB}/a_{BX}<4$ and beyond this interval it shows a saturation trend, due to the amplified population of trap states, see also Appendix~\ref{Ap:Same_width}. The latter renders the response of the system more enhanced compared to the $w/a_{\rm{ho}}=0.78$ case, since a larger number of post-quench eigenstates contributes in the dynamics [see also Appendix~\ref{Ap:Same_width}]. 
However, considering an initial state with a width at $w/a_{\rm{ho}}=4.9$, the response of the system is substantially enhanced as compared to the previous case and in particular it is almost independent of $a_{BB}/a_{BX}$. 
This pattern, as will be explained in ~\cref{Sec:LLH_small_width}, originates from the significant population of trap states.
In this sense, it becomes evident that there are two characteristic response regimes of the system with respect to $a_{\rm{ho}}$.

A similar qualitative behavior of $\langle |F| \rangle$ occurs also for LLH FFX settings [\cref{Fig:Fid_Av_LLH} (b)] at $w/a_{\rm{ho}}<1$ or $w/a_{\rm{ho}}>1$. 
Notably, for $w/a_{\rm{ho}}=0.78$, $\langle |F| \rangle$ is almost constant in the region $1/a_{FX}>3$. Such a response can be also observed for other widths $w/a_{\rm{ho}}<1$, due to the participation of trap states for large $1/a_{FX}$. For an initial state with $w/a_{\rm{ho}}=1$, we observe that the response of the LLH FFX system is decreased for $1/a_{FX}>3$, meaning that the deviation from the initial state reduces progressively.
This mainly occurs due to the smaller number of contributing states in the course of the evolution (as thoroughly discussed in the Appendix A of Ref. \cite{Gorin_dynamics_2006}), since the participation of the first two atom-dimers reduces as $1/a_{FX}$ is further tuned to larger values [see also \cref{Ap:Same_width}]. For $w/a_{\rm{ho}}=4.9$, the time-averaged fidelity is practically constant due to the participation of trap states during the dynamics, whose overlap coefficients do not depend strongly on $1/a_{FX}$ [see also \cref{Sec:LLH_large_width}]. Note that the considered post-quench $1/a_{FX}$ values do not include 0, and $\langle \abs{F} \rangle$ therefore deviates from unity in the leftmost part in Fig. \ref{Fig:Fid_Av_LLH} (b) at $w/a_{\rm{ho}}=1$. However, when $w \neq a_{\rm{ho}}$, even at $1/a_{FX}=0$ the deviation would persist, since the initial state is not a non-interacting eigenstate.

Evidently, regardless of the particle statistics we observe that the width of the initial state plays a crucial role on the dynamical response of the three-body system. 
Thus, in order to further address the physical origin of this behavior in the following we will analyze the involved excitations, in terms of the post-quench eigenstates, that contribute in the nonequilibrium dynamics. 
Their identification is indeed, in general, tractable in few-body setups~\cite{mistakidis2014interaction,mistakidis2017mode}.
For this purpose, we utilize the fidelity spectrum 
\begin{equation}
\abs{F(\tilde{\omega})}=\abs{\int \frac{dt}{\sqrt{2\pi}} \,e^{-i \tilde{\omega} t} \abs{F(t)}}.
\label{Eq:Fid_Spec}
\end{equation}
It discloses information regarding the predominantly contributing final eigenstates in the dynamics via the energy differences $\tilde{\omega}_{\rm{f,f'}}=E_{\rm{f}}-E_{\rm{f}'}$ (recall that we work with dimensionless units [\cref{Sec:Hyper}]), which are identified from the energy spectra of BBX and FFX systems~\cite{Bougas_Few_2021}. 
Below, we elaborate on the excitation spectrum of both LLH BBX and FFX systems in the two above-mentioned distinct response regimes.

\subsection{Excitations from narrow initial states with $w<a_{\rm{ho}}$} \label{Sec:LLH_small_width}

As a prototype LLH setup with an initial state width $w<a_{\rm{ho}}$ we use the case of $w/a_{\rm{ho}}=0.78$. 
To understand the excitation processes of the quenched system we inspect the respective fidelity spectrum together with the energy eigenspectrum and the overlap coefficients. 
For the BBX system, the fidelity spectrum $\abs{F(\tilde{\omega})}$ and the three-body post-quench eigenenergies are shown in \cref{Fig:Fid_Spec_LLH} (a), and (b), respectively.  
Note that the indexing of the eigenenergies e.g. in \cref{Fig:Fid_Spec_LLH} (b) starts from the ground state, which possesses an energy way below the displayed range, and increases as we climb the energy ladder.

In \cref{Fig:Fid_Spec_LLH}, for $a_{BB}/a_{BX}<4$ the excited frequency branches appearing in $\abs{F(\tilde{\omega})}$ mainly refer to energy differences between the second trimer state (first excited trimer) $\rm{f}=2$, and either the first atom-dimer ($\rm{f}=3$) or the trap states ($\rm{f}=8$), see e.g $\tilde{\omega}_{3,2}$ and $\tilde{\omega}_{8,2}$ respectively in \cref{Fig:Fid_Spec_LLH} (a) at $a_{BB}/a_{BX}=2.5$. 
In these frequency branches the most dominant contribution in the coefficients $c_{\rm{f},\rm{in}}$ stems mainly from the second trimer.
This occurs since both the initial state and the second trimer are well localized at small values of the hyperradius, i.e. for $R<a_{\rm{ho}}$, yielding thus a large overlap.
In particular, for the frequency $\tilde{\omega}_{3,2}$ we observe that it remains constant as the scattering length ratio $a_{BB}/a_{BX}$ varies. 
This arises from the fact that the scattering length dependence of the second trimer and first atom-dimer eigenenergies is similar as shown in \cref{Fig:Fid_Spec_LLH}(b), thus their energy difference results into an almost constant frequency $\tilde{\omega}$.

As $a_{BB}/a_{BX}$ is tuned to larger values, the spatial extent of the post-quench eigenstates changes drastically [\cref{Fig:Fid_Spec_LLH} (b)], thus affecting their overlap with the initial configuration. 
Indeed, the participation of the second trimer state ($\rm{f}=2$) decreases for $a_{BB}/a_{BX}>4$. 
For these scattering length ratios the trimer and the atom-dimer states become tightly bound [see \cref{Fig:Fid_Spec_LLH} (b)]. 
Accordinly, their wave functions are much narrower than the initial one, which reduces the corresponding overlap coefficients.
In return, this results in a smaller amplitude of $\tilde{\omega}_{3,2}$, see \cref{Fig:Fid_Spec_LLH} (c) at $a_{BB}/a_{BX}=4.25$. 
This reduced contribution in the fidelity spectrum is counterbalanced by the enhanced population of more trap states giving rise to excitation branches whose values increase with larger $a_{BB}/a_{BX}$, see e.g. the scaling of $\tilde{\omega}_{8,2}$ in \cref{Fig:Fid_Spec_LLH} (a)] \footnote{Note that even if the labels of the post-quench eigenstates are the same, the frequency associated to them, $\tilde{\omega}_{8,2}$, acquires different values depending on the scattering length [\cref{Fig:Fid_Spec_LLH} (a)], since the energy spectrum changes drastically with respect to $a_{BB}/a_{BX}$.}. 
Their increasing behavior reflects the growing energy difference between the second trimer and trap states for $a_{BB}/a_{BX}>3$ [\cref{Fig:Fid_Spec_LLH} (b)]. 
Also, the amplitude $\tilde{\omega}_{8,2}$ increases with $a_{BB}/a_{BX}$ since the substantial spatial extent of the trap wave functions yields larger overlap with the initial state.
Furthermore, a larger number of branches arises in the fidelity spectrum as can be seen by comparing the profiles of $\abs{F(\tilde{\omega})}$ at $a_{BB}/a_{BX}=4.25$ and $a_{BB}/a_{BX}=2.5$ illustrated in \cref{Fig:Fid_Spec_LLH} (c). 
As a result, the response of the time-averaged fidelity for $w/a_{\rm{ho}}=0.78$ is more enhanced (smaller value of $\langle |F| \rangle$) for larger ratios of $a_{BB}/a_{BX}$ [\cref{Fig:Fid_Av_LLH} (a)]. Let us remark that time-dependent variation protocols of the scattering lengths would be of great interest, since they could result in a significant population of trimer states, even at the regimes where trap states acquire a large contribution.

\begin{figure}[t!]
\centering
\includegraphics[width=0.4\textwidth]{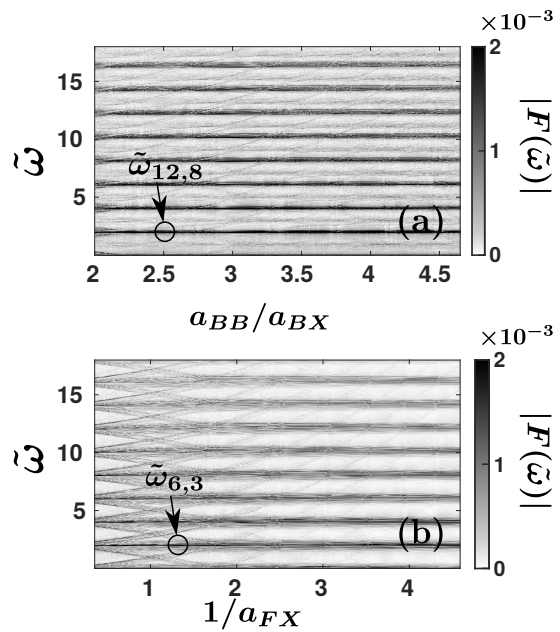}
\caption{Fidelity spectra of a wide initial state, i.e. $w/a_{\rm{ho}}=4.9$, for the (a) BBX and (b) FFX LLH systems following a quench of the scattering length. The circles designate specific frequency peaks corresponding to different post-quench eigenstates. The atoms reside in a superposition consisting predominantly of trap states. The latter are imprinted as excitation branches being insensitive to scattering length variations.}
\label{Fig:Fid_Spec_LLH_w_5}
\end{figure}

For the dynamical response of the LLH FFX system, we observe the appearance of a larger number of excitations in the fidelity spectrum [\cref{Fig:Fid_Spec_LLH} (d)] as $1/a_{FX}$ increases. 
Notice that this behavior is already anticipated from the enhanced response of $\langle |F| \rangle$ presented in \cref{Fig:Fid_Av_LLH} (b) for $w/a_{\rm{ho}}=0.78$. 
However, the microscopic mechanisms behind this response are different from the ones in the BBX system due to the distinct eigenenergy spectra, compare in particular \cref{Fig:Fid_Spec_LLH} (b) and (e). 
Evidently, in the case of the LLH FFX system trimers do not form. 
Here, the major contribution for $1/a_{FX}<2$ is shared among the first two atom-dimer states, $\rm{f}=1,2$, possessing a small spatial extent and mostly localized at $R<a_{\rm{ho}}$.  
This claim can be verified by the corresponding frequency peak $\tilde{\omega}_{2,1}$ of $\abs{F(\tilde{\omega})}$ shown in \cref{Fig:Fid_Spec_LLH} (d) and (f) as well as the contribution of the relevant overlap coefficients (with total contribution $90\%-60\%$ for $1/a_{FX} \in [0.36,2]$). 
For large scattering lengths ($1/a_{FX}>2$) the participation of atom-dimers diminishes since their spatial extent further decreases. 
This results in their reduced overlap with the initial state and consequently to a smaller amplitude of $\tilde{\omega}_{2,1}$ as shown in \cref{Fig:Fid_Spec_LLH} (f) for $1/a_{FX}=4$. 
In this case, trap states acquire a non-negligible population leading to interaction-dependent frequency branches which grow with respect to $1/a_{FX}$, see e.g. $\tilde{\omega}_{8,1}$ in \cref{Fig:Fid_Spec_LLH} (d).

\subsection{Response for wide initial configurations of $w>a_{\rm{ho}}$}  \label{Sec:LLH_large_width}

Next, we examine the susceptibility of LLH three-body setups to quenches for initial configurations characterized by $w>a_{\rm{ho}}$. 
As a representative example of this kind we choose $w/a_{\rm{ho}}=4.9$ and first investigate BBX mixtures.  
Recall that in this scenario the time-averaged response captured by $\langle |F| \rangle$ [\cref{Fig:Fid_Av_LLH} (a)] is drastically enhanced as compared to $w/a_{\rm{ho}}=0.78$ and experiences small variations with respect to $a_{BB}/a_{BX}$. 

To determine the microscopic origin of the involved excitations we resort again to the fidelity spectrum $\abs{F(\tilde{\omega})}$ provided in \cref{Fig:Fid_Spec_LLH_w_5} (a). 
The almost horizontal frequency branches stem from energy differences between trap states, e.g. $\tilde{\omega}_{12,8}$. This is verified by calculating the respective overlap coefficients and monitoring the energy spectrum [\cref{Fig:Fid_Spec_LLH} (b)]. 
Additionally, since $w/a_{\rm{ho}}=4.9 \gg 1$ the post-quench atom-dimers and trimers, being naturally narrow exhibit a reduced overlap with the initial state. 
The dominant contribution in the course of the evolution originates from the trap states whose overlap with $\Psi(R,\boldsymbol{\Omega},t=0)$ is appreciable. 
Indeed, a multitude of trap states is populated as can be inferred from the several frequency peaks of comparable amplitude appearing in $\abs{F(\tilde{\omega})}$ [\cref{Fig:Fid_Spec_LLH_w_5} (a)]. 
This fact, in turn, induces the enhanced response identified in $\langle |F| \rangle$ [\cref{Fig:Fid_Av_LLH} (a)] for $w/a_{\rm{ho}}=4.9$. 

A similar overall phenomenology takes place also for LLH FFX systems, see \cref{Fig:Fid_Spec_LLH_w_5} (b). 
Evidently, also here the respective excitation branches are almost insensitive to $1/a_{FX}$ variations [\cref{Fig:Fid_Spec_LLH_w_5} (b)]. 
Notably, the post-quench eigenstates responsible for this behavior are again trap states, e.g. $\tilde{\omega}_{6,3}$, although they are not the same as those identified in the BBX scenario [\cref{Fig:Fid_Spec_LLH_w_5} (a)]. 
The reason for this change can be traced back to the different structure of the eigenspectrum between BBX and FFX LLH systems, compare  Figs.~\ref{Fig:Fid_Spec_LLH} (b) and (e).

Focusing on the underlying selection processes according to which specific post-quench eigenstates are populated, it is instructive to carefully study the respective overlap coefficients. 
Of immediate interest here are the ones referring to pairs of post-quench eigenstates that experience avoided-crossings [dashed circles in \cref{Fig:Fid_Spec_LLH} (b), (e)], namely atom-dimers and trap states, and in particular illuminate their dependence on the width $w$ of the initial states.
In the vicinity of the avoided-crossings, the spatial extent of the involved eigenstates changes abruptly, since their character alters between trap and atom-dimer states. 

\begin{figure}[t!]
\centering
\includegraphics[width=0.4\textwidth]{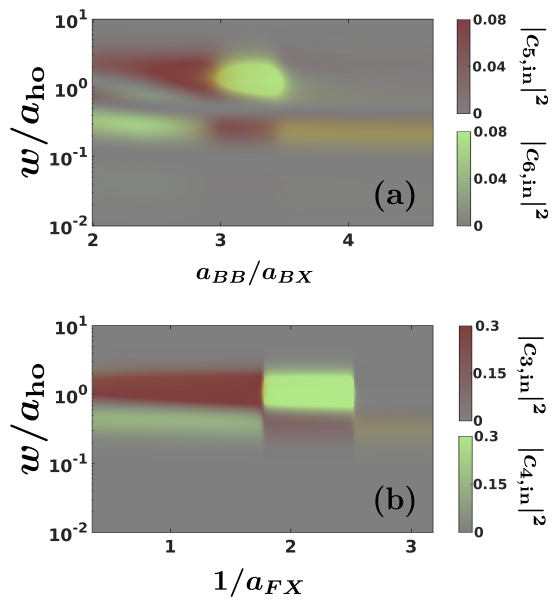}
\caption{Overlap coefficients $|c_{\rm{f},\rm{in}}|^2$ between the initial state of width $w$ and two post-quench eigenstates $\rm{f}$ as a function of $w/a_{\rm{ho}}$ and the scattering length for the (a) BBX ($\rm{f}=5,6$) and (b) FFX ($\rm{f}=3,4$) LLH system. In each case the presented pair of post-quench eigenstates experiences an avoided-crossing in the respective energy spectra [\cref{Fig:Fid_Spec_LLH} (b), (e) with dashed circles]. A change of the character of the state 
from a trap to an atom-dimer (atom-dimer to trap) state is signified by a shift of its major contribution to smaller (larger) values of the width of the pre-quench state $w/a_{\rm{ho}}$.}
\label{Fig:Overlaps_LLH}
\end{figure}

For a BBX setup, a characteristic example regarding the dependence of the overlap coefficients between the initial state and the $\rm{f}=5,6$ eigenstates as a function of $w$ and $a_{BB}/a_{BX}$ is displayed in \cref{Fig:Overlaps_LLH} (a).  
A transition between the different types of eigenstates is apparent by the complementary behavior of the respective overlap coefficients~\cite{dincao_efimov_2018}. 
On the left of the first avoided-crossing shown in \cref{Fig:Fid_Spec_LLH} (b) at $a_{BB}/a_{BX} \simeq 3$ [dashed circle], the occupation of the trap state $\rm{f}=5$ [see red color gradient in \cref{Fig:Overlaps_LLH} (a)] prevails for a larger $w$ when compared to the atom-dimer $\rm{f}=6$ [see green color gradient in \cref{Fig:Overlaps_LLH} (a)]. 
This behavior arises from the mere fact that the atom-dimer has a smaller spatial extent compared to the trap state, thus the latter yields larger overlap compared to the former. 
The opposite behavior takes place within $a_{BB}/a_{BX} \in [3, 3.5]$, since then the $\rm{f}=5,6$ states interchange their character. 
After the second avoided-crossing at $a_{BB}/a_{BX} \simeq 3.5$, these states are substantially occupied only for $0.19 < w/a_{\rm{ho}} < 0.39$, since then both of them are atom-dimers [\cref{Fig:Fid_Spec_LLH} (b)]. 
For larger $w/a_{\rm{ho}} \simeq 1$ and around $a_{BB}/a_{BX} \in [3.5,4]$ a significant contribution stems from a trap eigenstate  ($\rm{f}=7$), not shown in \cref{Fig:Overlaps_LLH} (a). 
Similar transitions occur also for the FFX LLH system [\cref{Fig:Overlaps_LLH} (b)], where in this case the pair of eigenstates $\rm{f}=3,4$ exchange character from a trap to an atom-dimer and vice versa through the avoided-crossing at $1/a_{FX} \simeq 1.77$ [\cref{Fig:Fid_Spec_LLH} (e) designated with a dashed circle]. 

\begin{figure*}[t]
    \centering
    \includegraphics[width=1\textwidth]{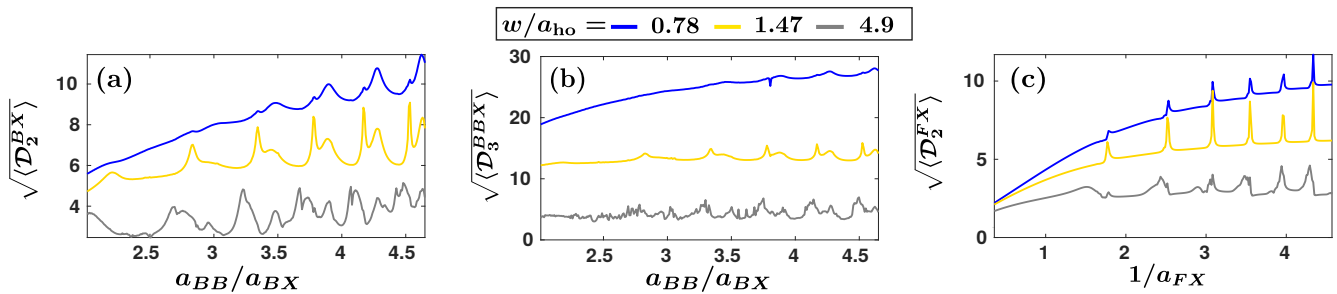}
    \caption{ Time-averaged (a) two-body
    $\sqrt{{\langle \mathcal{D}_2^{BX}} \rangle}$, and (b) three-body contact $\sqrt{\langle \mathcal{D}_3^{BBX} \rangle}$ of the BBX LLH setting and (c) two-body contact $\sqrt{{\langle \mathcal{D}_2^{FX}} \rangle}$ of the FFX setup. 
    Correlations at all levels increase for larger $a_{BB}/a_{BX}$ or $1/a_{FX}$ and their magnitude reduces for larger widths. 
    The peak structure at specific scattering lengths is an imprint of the participation of atom-dimers. 
    The widths of the initial state are provided in the legend.}
    \label{Fig:Avg_Cont2_Cont3_LLH}
\end{figure*}

Concluding, it is worth mentioning that upon considering a width of the initial state being the same as the three-body harmonic oscillator length, namely $w=a_{\rm{ho}}$, the original configuration corresponds to the non-interacting ground trap state [\cref{Sec:Setup}]. 
For this reason, the role of trimers and atom-dimers is less important during the time-evolution and as expected trap states have a somewhat larger population [for more details see Appendix~\ref{Ap:Same_width}]. 
This behavior holds for both BBX and FFX systems.

\subsection{Build-up of two- and three-body correlations} \label{Sec:Corr_LLH}

Having established an understanding regarding the contributing eigenstates for different widths of the initial state, an intriguing question that arises is how these states influence the associated short-range few-body correlations in the course of the evolution. 
These correlations can be addressed by the experimentally measurable~\cite{sagi_measurement_2012,stewart_verification_2010} 
two- and three-body contacts~\cite{olshanii_short-distance_2003,tan_large_2008,werner_general_2012,colussi_two-body_2019,bellotti_dimensional_2013,Bougas_Few_2021}. 
The latter are defined as coefficients in a high momentum expansion of the $\sigma$-species one-body density in momentum space 

\begin{equation}
n_{\sigma}(\boldsymbol{p}_{\sigma},t)\simeq \frac{1}{N_{\sigma}p_{\sigma}^4}\sum_{\sigma'} (1+\delta_{\sigma\sigma'})\mathcal{D}_2^{\sigma\sigma'}(t)+\frac{\ln^3 p_{\sigma}}{p^6_{\sigma}} \mathcal{D}_3(t).
\label{Eq:Contacts}
\end{equation}
This expansion pertains to the case where $p_{\sigma}$ is significantly larger than the momentum scales provided by the inverse scattering lengths~\cite{Bougas_Few_2021}. 
Here, $N_{\sigma}$ is the atom number belonging to the $\sigma$-species, while $\mathcal{D}_2^{\sigma\sigma'}(t)$ denotes the time-dependent two-body contact between the species $\sigma$ and $\sigma'$. 
Note that only the three-body contact $\mathcal{D}_3(t)$ of BBX systems ($\mathcal{D}_3^{BBX}(t)$) is finite, since for FFX ones three-body correlations are suppressed~\footnote{The three-body contact yields the probability to detect three particles in close vicinity. 
As such, it is zero by construction for FFX systems within the $s$-wave zero-range interaction model, where the two identical and non-interacting fermions can not approach one another due to the Pauli principle.} due to the Pauli exclusion principle~\cite{bellotti_contact_2014}. 
The main features of these few-body correlation observables are captured by their time-averaged measure. 
Namely, the time-averaged two-body contacts are described by the following expressions:

\begin{equation}
\langle \mathcal{D}_2^{\sigma X} \rangle=\lim_{T \to \infty} \frac{1}{T}\int_0^T dt \: \mathcal{D}_2^{\sigma X}(t), \quad \sigma=B, F
\label{Eq:Averaged_contacts2b}
\end{equation}
and the three-body ones read
\begin{equation}
\langle \mathcal{D}_3^{BBX} \rangle  =  \lim_{T \to \infty} \frac{1}{T}\int_0^T dt \: \mathcal{D}_3^{BBX}(t).
\label{Eq:Averaged_contacts3b}
\end{equation}

These quantities assess the overall degree of dynamical correlations for various widths of the initial state and post-quench scattering lengths, see \cref{Fig:Avg_Cont2_Cont3_LLH}. 
A detailed analysis of the stationary three-body FFX and BBX setups reveals a hierarchy in terms of the degree of few-body correlations for the different types of eigenstates. 
Namely, as shown in Refs.~\cite{Bougas_Few_2021,colussi_two-body_2019,Blume_harmonically_2018} trimer states possess more enhanced two- and three-body correlations than those of the BX or FX atom-dimer states and, similarly, the atom-dimer contacts are larger than those of the trap states.
This hierarchy will also be apparent here as the width of the initial state changes and different eigenstates contribute in the dynamical response.
Indeed, as the width of the initial state [Eq.~\eqref{Eq:Non_int_Gauss_Hyper}] increases, the magnitude of all the aforementioned correlations at any scattering length is reduced [\cref{Fig:Avg_Cont2_Cont3_LLH}].
This occurs because for larger widths, a superposition of trap states is predominantly populated [see also \cref{Fig:Fid_Spec_LLH_w_5}].

On the contrary, for $w/a_{\rm{ho}}=0.78$, the first two atom-dimers (second trimer) provide the main contribution to the post-quench wave function Eq.~\eqref{Eq:Quench} of the FFX (BBX) system. This is confirmed through their dominant 
overlap coefficients [see \cref{Sec:LLH_small_width}], enhancing few-body correlations compared to cases where $w>a_{\rm{ho}}$ [\cref{Fig:Avg_Cont2_Cont3_LLH}]. 
Therefore, in the limit of small $w<a_{\rm{ho}}$, correlations at the two- and three-body level are, generically, enhanced due to the non-negligible involvement of trimer and atom-dimer states.
This amplification was also observed for a three-boson setup in the quench dynamics at unitarity in 3D~\cite{colussi_dynamics_2018}, especially when the width of the initial state matched the size of an Efimov trimer. 

Another remarkable feature of the correlations is their magnification at particular scattering lengths for fixed $w$, see the individual peaks displayed in \cref{Fig:Avg_Cont2_Cont3_LLH}. 
Their amplitudes become more prominent from the overall two- and three-body contacts for increasing width $w$, where trap states contribute substantially [see \cref{Sec:LLH_large_width}]. 
These peaks occur in the vicinity of avoided-crossings present in \cref{Fig:Fid_Spec_LLH} (b), (e) where the corresponding three-body wave function is predominantly in a superposition of a trap and an atom-dimer state.
Therefore, in this range of scattering length ratios the overall character of the wave function abruptly changes yielding in this manner an enhanced $\langle \mathcal{D}_2^{BX} \rangle$, $\langle \mathcal{D}_3^{BBX} \rangle$ and $\langle \mathcal{D}_2^{FX} \rangle$.
This particular property of the time-averaged two- and three-body contacts can be utilized as an experimental probe for the formation of atom-dimers in a 2D gas.

Furthermore, the appearance of enhanced peaks in the two-body contacts at the avoided-crossings due to the atom-dimer component in the time-evolved wave function is also a manifestation of the universal Tan relations. 
These universal relations exemplify that the short-range two-body correlations are proportional to the variation of the stationary energy spectra with respect to the scattering length~\cite{werner_general_2012,valiente_universal_2011}. 
Therefore, close to the avoided-crossings the eigenenergies of the three-body system [see Figs.~\ref{Fig:Fid_Spec_LLH} (b), (e)] strongly vary with the scattering length thus yielding narrow peaked two-body correlations [Figs.~\ref{Fig:Avg_Cont2_Cont3_LLH} (a), (c)]. 
By this token, we can address the main difference between the two-body contacts of BBX and FFX systems in \cref{Fig:Avg_Cont2_Cont3_LLH}(a) and (c), respectively, where the former exhibits broader peaks than the latter.
This occurs because in the FFX eigenspectra shown in \cref{Fig:Fid_Spec_LLH}(e) we observe much sharper avoiding-crossings than in the BBX ones [see \cref{Fig:Fid_Spec_LLH}(b)].
Such a universal relation is absent in the case of the three-body contact~\cite{bellotti_dimensional_2013} in 2D, and the peak structure is attributed to the enhanced stationary three-body correlations~\cite{Bougas_Few_2021} of the atom-dimer component of the time-evolved wave function.

\begin{figure}[t!]
\centering
\includegraphics[width=0.4 \textwidth]{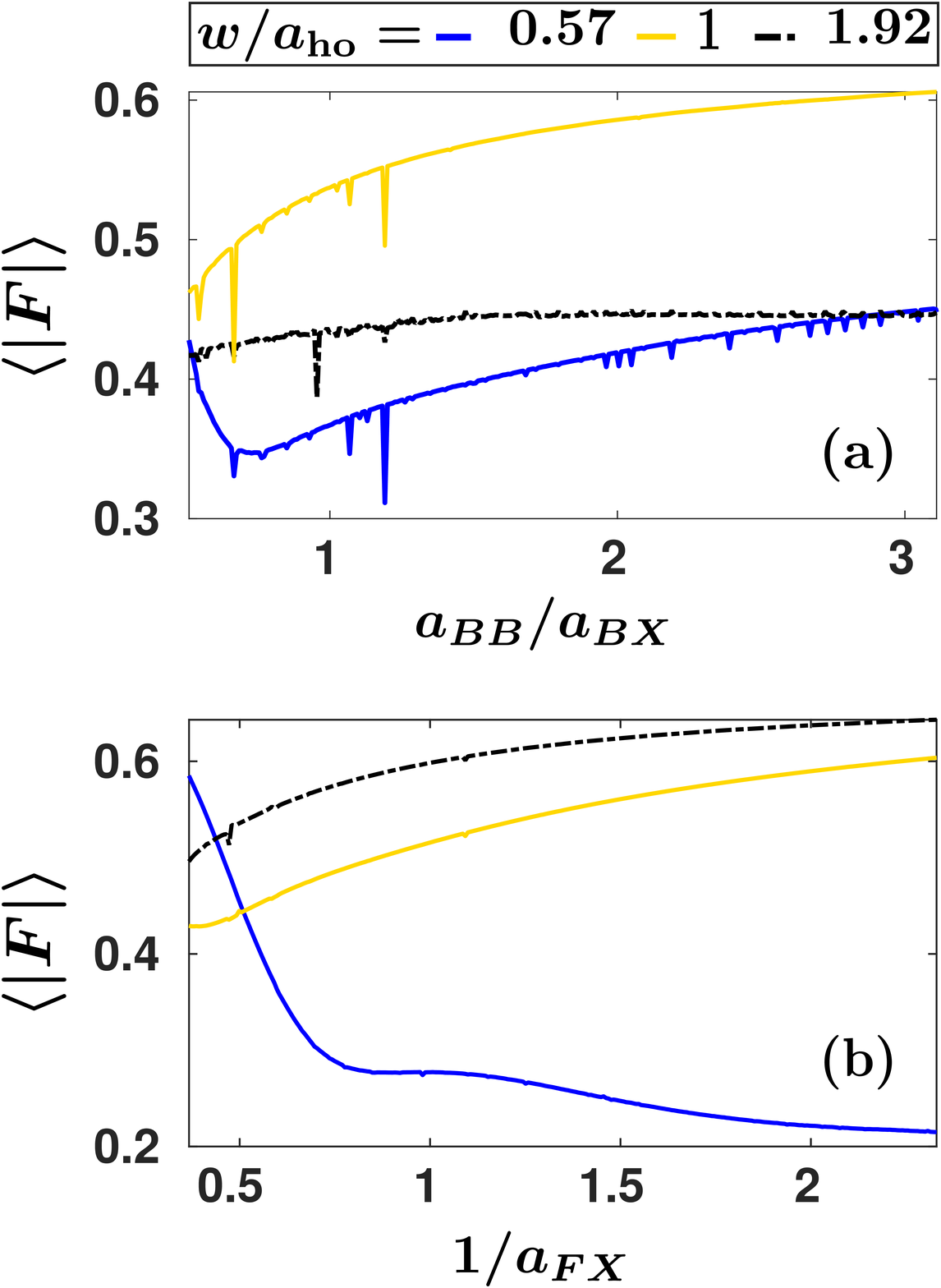}
\caption{Time-averaged fidelity $\langle \abs{F} \rangle$, of the three-body (a) BBX and (b) FFX HHL mixture subjected to quenches of the interspecies scattering length. 
Different widths of the initial state are considered (see legend) whose values in terms of the oscillator length ($a_{\rm{ho}}=2.6$) determine the degree of the system's response.   
The substantial population of atom-dimer and trimer (trap) states for $w<a_{\rm{ho}}$ ($w>a_{\rm{ho}}$) leads to a strongly (weakly) interaction dependent response. In contrast to the LLH case, trap states have also a small contribution for $w<a_{\rm{ho}}$ in addition to trimers and atom-dimers, and the larger number of participating eigenstates compared to the $w>a_{\rm{ho}}$ scenario, enhances the response of the system.} 
\label{Fig:Fid_Av_HHL}
\end{figure}

Moreover, it is also worth mentioning that a broadening of these correlation peaks is evident for larger widths, see e.g. $w/a_{\rm{ho}}=4.9$ in \cref{Fig:Avg_Cont2_Cont3_LLH}. 
In this case, as already discussed and observed in the fidelity spectrum [\cref{Fig:Fid_Spec_LLH_w_5}] a large amount of trap states participates in the three-body time-evolved wave function.  
This results into an agglomeration of avoided-crossings contributing to the dynamics, which are slightly displaced horizontally from one another at a fixed scattering length [see \cref{Fig:Fid_Spec_LLH} (b), (e)]. 
The aforementioned displacement then yields a range of scattering lengths over which the Tan contacts display an enhanced behavior, manifested as a peak broadening.

\section{Dynamical response of HHL mixtures}\label{Sec:Dyn_HHL}

In this section we address the role of the masses on the dynamical build up of few-body correlations by considering HHL three-body mixtures.
The intrinsic dynamical behavior of this system is explored, for widths $w$ of the initial state smaller or larger than the characteristic three-body harmonic oscillator length $a_{\rm{ho}}=2.6$ [Eq.~\eqref{Eq:Non_int_Gauss_Hyper}]. 
As in the LLH case in \cref{Sec:Dyn_LLH}, we remark that initial states with a spatial extent smaller (larger) than $a_{\rm{ho}}$ favors the participation of trimer and/or atom-dimer (trap) states. 
Our analysis on the response of the 2D mixtures is based on the time-averaged fidelity $\langle |F| \rangle$ given in Eq.~(\ref{Eq:Fid_av}). 

\begin{figure*}[t]
\centering
\includegraphics[width=1.0\textwidth,keepaspectratio]{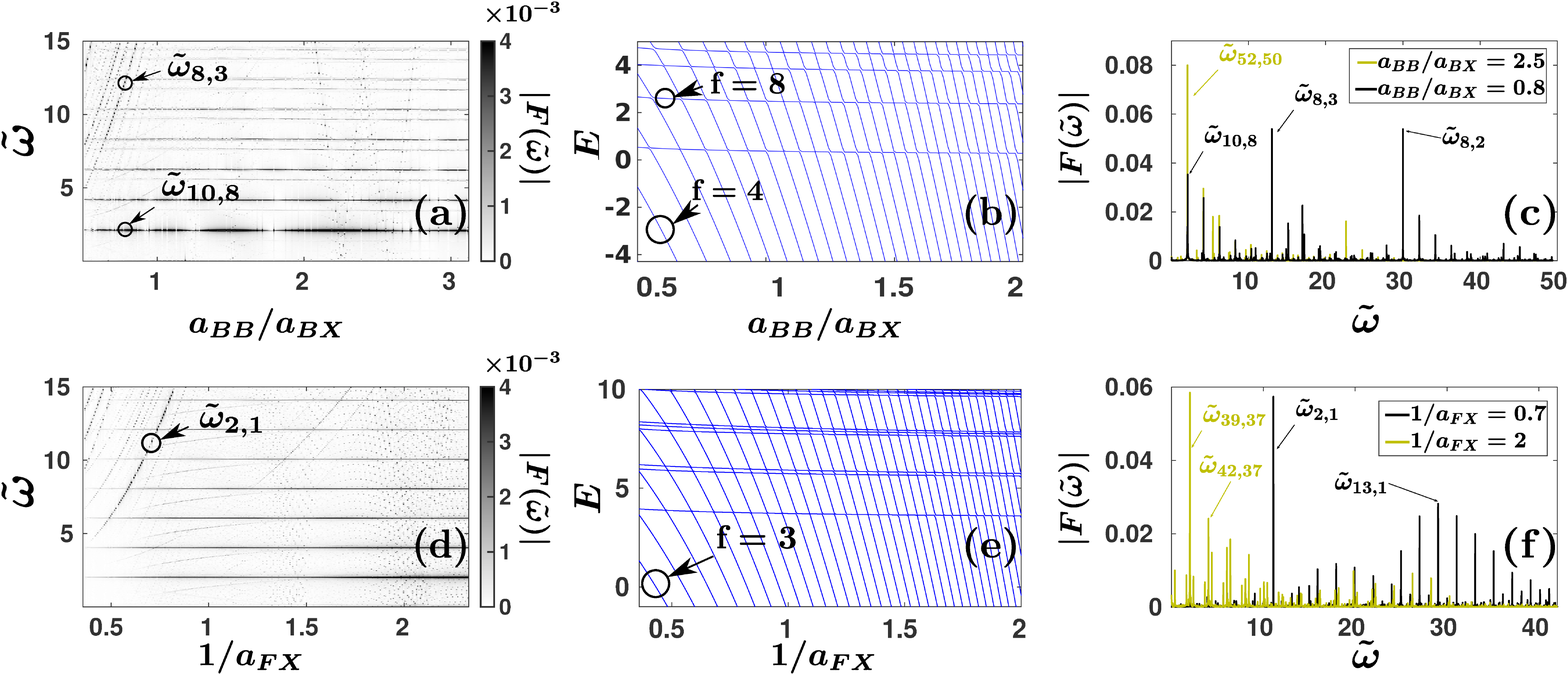}
\caption{Fidelity spectrum of the (a) BBX and (d) FFX HHL systems performing an interaction quench of an initial state where $w/a_{\rm{ho}}=0.57$. 
The arrows mark characteristic frequency branches $\tilde{\omega}_{f,f'}$. 
Excitation branches that alter with respect to the scattering length (see the top left corners) correspond to energy differences between trimers, atom-dimers and trap states. 
Otherwise, the almost fixed frequency branches refer to trap states. 
Energy spectra of (b) BBX and (e) FFX HHL mixtures.  
Particular eigenstates are denoted by circles and arrows. 
Specific profiles of the fidelity spectrum of the (c) BBX and (f) FFX system at distinct scattering lengths (see legends).}
\label{Fig:Fid_Spec_HHL}
\end{figure*}

The overall response of a HHL BBX system characterized by $m_B/m_X=22.16$ is intensified in the case of $w/a_{\rm{ho}}=0.57$ [\cref{Fig:Fid_Av_HHL} (a)] as compared to $w>a_{\rm{ho}}$ within $a_{BB}/a_{BX} \in [0.5,3]$. 
This is in contrast to the susceptibility of LLH mixtures [\cref{Fig:Fid_Av_LLH} (a)]. 
Moreover, for $w/a_{\rm{ho}}=1$ a strong dependence of $\langle \abs{F} \rangle$ is observed with respect to the scattering length ratio. 
This feature of $\langle \abs{F} \rangle$ differs dramatically from the response for $w/a_{\rm{ho}}=1.92$, where it 
is arguably almost insensitive within the interaction interval $a_{BB}/a_{BX} \in [2,3]$. 
This behavior is related to the prominent contribution of trap states.
For $w/a_{\rm{ho}}=1$, the system becomes less susceptible to the quench as compared to the case of $w/a_{\rm{ho}}=1.92$, since fewer trap states contribute, especially for large $a_{BB}/a_{BX}>2$. 
Notably, there is a series of peaks appearing in $\langle \abs{F} \rangle$ at specific scattering lengths, where avoided-crossings among atom-dimer and trap states exist in the few-body eigenspectrum [see also \cref{Fig:Fid_Spec_HHL} (b), (e)]. 
Their importance, especially in the relevant few-body correlations, will be discussed below. 

Subsequently, the susceptibility of a HHL FFX system with $m_F/m_X=24.71$ is illustrated in \cref{Fig:Fid_Av_HHL} (b). 
Apparently, the FFX mixture becomes more perturbed when considering $w/a_{\rm{ho}}=0.57$. 
For larger widths, e.g. $w/a_{\rm{ho}}=1.92$, the system experiences a weak dependence on the scattering length within the range $1/a_{FX} \in [1.5,2]$. 
This is linked to the dominant presence of trap states during the time-evolution due to their large spatial extent. 
Moreover, we note that similarly to the BBX HHL case [\cref{Fig:Fid_Av_HHL} (a)] the FFX mixture is less perturbed for $w/a_{\rm{ho}}=1.92$ 
than in the $w/a_{\rm{ho}}=0.57$ scenario. 
However, in contrast to the HHL BBX system, for $w/a_{\rm{ho}}=1$ the mixture develops a stronger response in comparison to $w/a_{\rm{ho}}=1.92$, due to the more prominent population of trimers and atom-dimers.

\subsection{Excitation processes for $w<a_{\rm{ho}}$}  \label{Sec:HHL_small_width}

Pre-quenched states with a spatial extent smaller than the three-body harmonic oscillator length, apparently exhibit a larger overlap with the trimers and atom-dimer states of the BBX and FFX HHL systems. 
The latter, contribute significantly in the underlying dynamics compared to the case where $w>a_{\rm{ho}}$. In the opposite regime ($w>a_{\rm{ho}}$) trap states become substantially populated in the post-quench dynamics [see also \cref{Sec:LLH_large_width}], a mechanism pertaining also to the HHL mixtures. The frequency spectra will be analyzed for the $w<a_{\rm{ho}}$ scenario, since for $w>a_{\rm{ho}}$, the underlying microscopic mechanisms resemble those presented in \cref{Sec:LLH_large_width}. However, the differences present in $\langle \abs{F} \rangle$ between LLH [Fig. \ref{Fig:Fid_Av_LLH}] and HHL setups [Fig. \ref{Fig:Fid_Av_HHL}] for $w>a_{\rm{ho}}$ stem mostly from the different number of participating trap states in the post-quench dynamics. Moreover, in the HHL scenario, in addition to the participation of trap states, there are a few contributing atom-dimer and trimer states especially for small values of $1/a_{FX}$ and $a_{BB}/a_{BX}$. This results in further perturbation of the system from the initial state compared to larger scattering lengths. Recall also here the relevant discussion in \cref{Sec:Dyn_LLH} concerning LLH mixtures. 

\begin{figure*}[t]
\centering
\includegraphics[width=1 \textwidth]{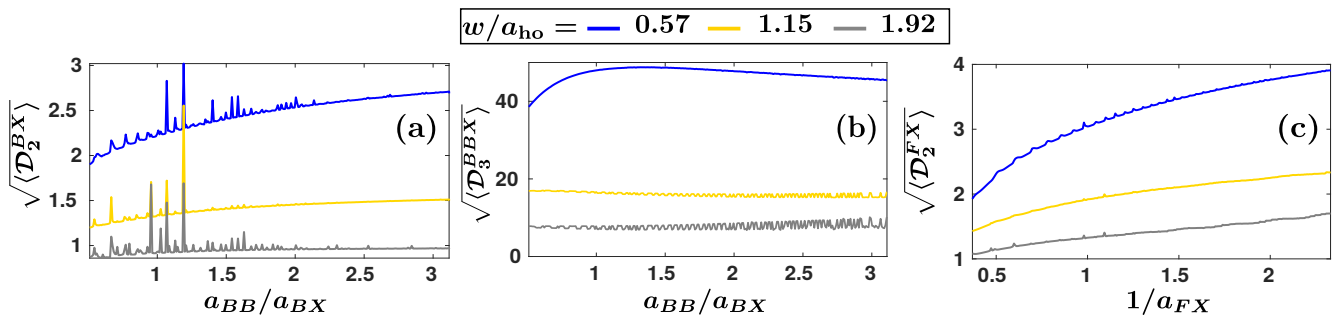}
\caption{Time-averaged contacts revealing the enhancement of short range (a), (c) two-body and (b) three-body correlations for larger inverse interspecies scattering lengths of HHL (a), (b) BBX and (c) FFX settings. 
The existence of peaks at individual scattering length ratios reveals the population of atom-dimers due to the sharp avoided-crossing taking place at the eigenspectrum [\cref{Fig:Fid_Spec_HHL} (b), (e)].  
The widths of the initial state are shown in the legend.}
\label{Fig:Avg_Cont2_Cont3_HHL}
\end{figure*}

Inspecting the fidelity spectrum $\abs{F(\tilde{\omega})}$ [\cref{Fig:Fid_Spec_HHL} (a)] together with the overlap coefficients and the energy spectrum [\cref{Fig:Fid_Spec_HHL} (b)] for the HHL BBX system, we can infer that for $a_{BB}/a_{BX}<1$ the second and third trimer states are significantly populated.  
This gives rise to excitation branches such as $\tilde{\omega}_{8,3}$, indicating the participation of the third trimer and the $\rm{f}=8$ trap state [\cref{Fig:Fid_Spec_HHL} (a)] for small $a_{BB}/a_{BX}<1$. 
This frequency branch shows an appreciable growth with larger $a_{BB}/a_{BX}$ due to the accompanied increasing energy difference between trimer and trap states [\cref{Fig:Fid_Spec_HHL} (a), (b)]. 
Note that the energies of the $\rm{f}=2,~3$ trimers are large in magnitude and negative and therefore lie below the energy window presented in \cref{Fig:Fid_Spec_HHL} (b). 
Apart from trimer states, trap ones, e.g. $\rm{f}=8,~10$, are occupied as well but their respective energy differences depend weakly on changes of 
$a_{BB}/a_{BX}$, see e.g. $\tilde{\omega}_{10,8}$ in \cref{Fig:Fid_Spec_HHL} (a).

A further increase of the scattering length ratio $a_{BB}/a_{BX}>1$, leads to a reduction of the amplitude and number of the higher-lying excitation frequencies in comparison to $a_{BB}/a_{BX}<1$. 
This behavior can be readily seen in the relevant profiles of the fidelity spectra depicted in \cref{Fig:Fid_Spec_HHL} (c) for $a_{BB}/a_{BX}=2.5$ and $a_{BB}/a_{BX}=0.8$. 
It stems from the suppressed contribution of the two trimer states for $a_{BB}/a_{BX}>1$, resulting in a less perturbed system as also reflected in $\langle \abs{F} \rangle$ [\cref{Fig:Fid_Av_HHL} (a)] for $w=1.5$.  
Similarly to the case of $a_{BB}/a_{BX}<1$ trap states are also populated here, imprinted in the spectrum as distinct almost horizontal  frequency branches e.g. $\tilde{\omega}_{52,50}$ in \cref{Fig:Fid_Spec_HHL} (c) 
\footnote{Apart from the horizontal excitation branches within $a_{BB}/a_{BX} \in [2,3]$, there exist also faint ones 
having a 'V' shape dependence on the scattering length with tipping points located at $a_{BB}/a_{BX}=2.25$ and $2.57$ [\cref{Fig:Fid_Spec_HHL} (a)]. These are attributed to energy differences between trap and atom-dimer states. At the tipping point of these 'V' shaped branches the energies of participating states come close together due to the avoided-crossings and are thus associated to small $\tilde{\omega}$ in $\abs{F(\tilde{\omega})}$.}.

A qualitatively similar dynamical response to the BBX mixture is also observed for the HHL FFX system, see $\abs{F(\tilde{\omega})}$ illustrated in \cref{Fig:Fid_Spec_HHL} (d) for $w/a_{\rm{ho}}=0.57$. 
Here, the heavy fermions with respect to the third particle favor trimer formation~\cite{pricoupenko_universal_2010},  
a result that is in contrast to the corresponding LLH case. 
These trimer states possess large negative energies~\cite{Bougas_Few_2021,bellotti_mass-imbalanced_2013}, lying beyond the values depicted in the energy spectrum provided in \cref{Fig:Fid_Spec_HHL} (e). 
Particularly, a superposition of the first two trimer states ($\rm{f}=1$ and $2$) is prevalent in the course of the evolution for $1/a_{FX}<1$, leading to excitation branches such as $\tilde{\omega}_{2,1}$ [\cref{Fig:Fid_Spec_HHL} (d)]. 
Moreover, similar to the BBX HHL system, trap states are also present in the dynamical response of the corresponding FFX mixture, as identified by the energy spectrum and the overlap coefficients. The frequency branches associated to energy differences between these states are almost independent of $1/a_{FX}$ [\cref{Fig:Fid_Spec_HHL} (d)].

Tuning the inverse scattering length to larger values $1/a_{FX}>1$, a plethora of trap states contributes in the time-evolved three-body wave function. 
Accordingly, a multitude of excitation branches arise in $\abs{F(\tilde{\omega})}$ whose location is almost constant with varying $1/a_{FX}$ [\cref{Fig:Fid_Spec_HHL} (d)] and are clustering at low $\tilde{\omega}$ as shown in \cref{Fig:Fid_Spec_HHL} (f). 
The large number of contributing trap states for $1/a_{FX}>1$ is linked to the  enhanced response of the HHL FFX system, e.g. captured by the time-averaged fidelity displayed in  \cref{Fig:Fid_Av_HHL} (b) for $w/a_{\rm{ho}}=0.57$.

\subsection{Dynamical formation of few-body correlations}  \label{Sec:Corr_HHL}

As already demonstrated in \cref{Sec:Corr_LLH} for LLH systems, the build-up of few-body correlations regardless of the particle statistics exhibit a peak structure for scattering lengths in the vicinity of avoided-crossings appearing in the post-quench eigenspectrum (see also \cref{Fig:Avg_Cont2_Cont3_LLH}).
Similarly, in this section we focus on HHL systems in order to showcase the role of increased mass ratio on the time-averaged Tan contacts as illustrated in \cref{Fig:Avg_Cont2_Cont3_HHL}.
In particular, the two-body BX species contact [\cref{Fig:Avg_Cont2_Cont3_HHL} (a)] exhibits sequences of narrow peaks at specific scattering length ratios  in agreement with Tan's universal relation~\cite{tan_energetics_2008,werner_general_2012,braaten_universal_2011}.
Namely, at these post-quench scattering lengths, the corresponding eigenspectrum possesses narrow avoided-crossings among trap states and atom-dimers [\cref{Fig:Fid_Spec_HHL} (b)], thus resulting into the strong amplification of the two-body correlations. 
Moreover, the amplitude of the peaks in the $\braket{\mathcal{D}_2^{BX}}$ decreases for large $a_{BB}/a_{BX}$ independently of $w$. 
This suppression occurs for large $a_{BB}/a_{BX}$ where the avoided-crossings become increasingly narrow [\cref{Fig:Fid_Spec_HHL} (b)]. In this sense, they can not be well resolved leading to less pronounced peaks compared to smaller $a_{BB}/a_{BX}$. 

In the case of the three-body contact [\cref{Fig:Avg_Cont2_Cont3_HHL} (b)] a multitude of peaks with tiny amplitude appears as $w$ increases.
This holds even for large $a_{BB}/a_{BX}$ as $w$ increases, despite the narrow avoided-crossings present in the HHL eigenspectra [\cref{Fig:Fid_Spec_HHL} (b)]. 
Particularly, for increasing $w$ trap states are predominantly populated,
but in the vicinity of avoided-crossings atom-dimers contribute as well. 
Therefore, the amplification of stationary three-body correlations of the atom-dimer post-quench eigenstates compared to trap states leads to the rise of peaks in $\sqrt{\langle \mathcal{D}_3^{BBX} \rangle}$ at the locations of the avoided-crossings. 
Moreover, equivalently to the two-body BX species contact [\cref{Fig:Avg_Cont2_Cont3_HHL} (a)], the time-averaged three-body contact is reduced for larger $w$, due to the significant participation of trap states, whose stationary three-body correlations are greatly suppressed. 

Furthermore, $\sqrt{\langle \mathcal{D}_3^{BBX}\rangle}$ at $w/a_{\rm{ho}}=0.57$ has an overall maximum around $a_{BB}/a_{BX}\simeq 1.2$, and then decreases for larger values of $a_{BB}/a_{BX}$. 
This behavior is related to the significant population of the second trimer which specifically possesses a population up to $16\%$ until $a_{BB}/a_{BX} \simeq 1.2$.
Subsequently, the corresponding overlap coefficient with the initial state decreases for $a_{BB}/a_{BX}>1.2$, since in this range of scattering length ratios the second trimer state is narrower than the initial one.
HHL BBX systems favor the existence of strongly bound trimer states, due to the increased mass ratio~\cite{bellotti_mass-imbalanced_2013}. The contribution of such a trimer state (second) for $w<a_{\rm{ho}}$, results in an augmented three-body contact, in contrast to the one presented in LLH setups [\cref{Fig:Avg_Cont2_Cont3_LLH} (b), $w/a_{\rm{ho}}=0.78$], where the small mass ratio inhibits the creation of strongly bound trimers.

In an equal fashion to the time-averaged two-body BX contact, $\sqrt{\langle \mathcal{D}_2^{FX} \rangle}$ [\cref{Fig:Avg_Cont2_Cont3_HHL} (c)] showcases small amplitude peaks, arising mostly for $w/a_{\rm{ho}}=0.57$.
Their magnitude again drops for increasing scattering length ratio $1/a_{FX}$ since sharper avoided-crossings are encountered in the eigenspectrum of the HHL FFX system than the ones appearing in the LLH case [compare \cref{Fig:Fid_Spec_LLH} (e) and \cref{Fig:Fid_Spec_HHL} (e)].

\section{Experimental parameters for the realization of the dynamics of the three-body mixture} \label{Sec:Exp}

In an experimental environment 2D gases are realized in quasi-2D trapping potentials where the confinement in the transversal direction of the 2D plane is tighter than the radial one.
This transversal trapping component is characterized by a frequency $\omega_{\perp}$ chosen such that the atomic motion is energetically restricted to the radial confinement potential with frequency $\omega_r$ ~\cite{Petrov_interatomic_2001,He_s-wave_2019}. 
A comparison of the low-lying energy states of two interacting particles in 3D and in a pure 2D geometry~\cite{Idziaszek_Analytical_2006}, revealed that the aspect ratio in a quasi-2D setup required to attain the 2D character of the relative motion of the two particles~\cite{Petrov_interatomic_2001} should satisfy $\omega_r/\omega_{\perp}<1/10$. 
This is corroborated by typical quasi-2D experiments~\cite{Kwon_spontaneous_2021,holten_anomalous_2018,murthy_quantum_2019}. 
For our setup, this energy requirement translates to $1/(\mu w^2) \leq 0.1 \omega_{\perp}$, and furthermore assuming $\omega_{\perp}=50$~\cite{Makhalov_ground_2014}, it reduces to $w\geq 1/\sqrt{5\mu}$. 
As such, for the typical LLH settings that we have considered this condition yields $w \geq 0.4559$, while for HHL ones it yields $w \geq 1.16$. 

The dynamical protocol outlined in \cref{Sec:Setup} relies on the realization of a non-interacting three-body system with a tunable spatial extent $w$, and the subsequent quench of the relevant 2D scattering lengths. 
The latter are related to their 3D counterparts~\cite{Petrov_interatomic_2001}, which can be tuned by means of Feshbach resonances~\cite{chin_feshbach_2010}. 
For the BBX systems, in particular, the coexistence of broad and narrow intra- and interspecies resonances in a magnetic field window ensures a regime where the post-quench scattering length $a_{BB}$ remains almost constant while $a_{BX}$ varies in magnitude and sign. 
For instance, for the HHL BBX system of $^{133}\rm{Cs}-^{133}\rm{Cs}-^6\rm{Li}$, such a magnetic field window exists for $[840,845] \rm{G}$, i.e. around the interspecies resonance~\cite{berninger_feshbach_2013,pires_analyzing_2014,Repp_observation_2013}. 
Also, in the vicinity of $\simeq 880 \, \rm{G}$ both 3D scattering lengths vanish, thus materializing a non-interacting state. 

The parameters of interest for the trapping potential are $\omega_r=2 \pi \times 65 \, \rm{kHz}$ and $\omega_{\perp}=50 \,\omega_r$~\cite{Makhalov_ground_2014}.
Also, regarding the 3D counterparts of the 2D post-quench scattering lengths used herein, we discern the following values displayed in \cref{TAB:scattering}. Note that in the considered intervals of the 3D scattering length (in atomic units), there is a sign change due to a resonance.

Our analysis in the previous sections illustrated the role of the width $w$ of the initial state in the dynamical response of the three-body system.
This $w$ parameter can be experimentally adjusted by the following procedure.
The two identical particles (B or F) together with the third distinguishable atom (X) are confined in a trap with a planar frequency $\omega_{\rm{in}}$, which are initialized in their non-interacting ground state.
A simple relation can be established between the initial state's width and the planar frequency, i.e. $\mu \omega_{\rm{in}}= w^{-2}$, where $\mu$ is the three-body reduced mass [see also Sec.~\ref{Sec:Setup}]. 
Prior to the quench on the scattering lengths, a quench on the trap frequency from $\omega_{\rm{in}}$ to $\omega_{\rm{f}}$ is performed. 
This allows for the preparation of initial states that possess widths different from the length scale of the trap with final frequency $\omega_{\rm{f}}$ where the interaction quench dynamics will take place.
By setting the final radial trapping frequency at $\omega_{\rm{f}}=2\pi \times 65 \, \rm{kHz}$, the initial frequency is determined from the relation $\omega_{\rm{in}}=\omega_{\rm{f}}\, a_{\rm{ho}}^2/w^2$.
Thus, for the LLH settings in Sec.~\ref{Sec:Dyn_LLH}, the widths $w/a_{\rm{ho}}=0.78, ~ 4.9$ correspond to $\omega_{\rm{in}}=2\pi \times (105.5, 2.7) \, \rm{kHz}$.
For the HHL setup (Sec.~\ref{Sec:Dyn_HHL}), the initial widths $w/a_{\rm{ho}}=0.57, ~ 1.92$, are obtained for $\omega_{\rm{in}}=2\pi \times (194.5, 17.5) \, \rm{kHz}$. 

\begin{table}[t!]
    \centering
    \renewcommand{\arraystretch}{1.7}
    \setlength{\tabcolsep}{4pt}
    \begin{tabular}{|c|c|c|} \hline
         & $1/a_{FX}$ & $a^{3D}_{FX} \: (a_0)$  \\ \hhline{|=|=|=|}
      LLH   & $[0.36,2.77] \, ([4,5])$ & $[-246, -3000] \, ([3000,1343])$ \\ \hline
      HHL & $[0.36, 6]\, ([0.82, 2.5])$ & $[-715,-2976]\, ([2991,380])$ \\ \hhline{|=|=|=|}
       &  $1/a_{BX}$ & $a^{3D}_{BX} \: (a_0)$  \\ \hhline{|=|=|=|}
        LLH  &  $[2,2.81] \, ([3.94,4.65])$ & $[-994,3000] \, ([2995,1497])$ \\ \hline
        HHL & $[0.85,3]$ & $[3000,380]$ \\ \hhline{|=|=|=|}
        & $a_{BB}$ & $a^{3D}_{BB} \: (a_0)$ \\ \hhline{|=|=|=|}
        LLH & $1$ & $-421$ \\ \hline
        HHL & $1$ & $1578$ \\ \hline
    \end{tabular}
    \caption{Mapping of the 2D BX, FX and BB post-quench scattering lengths to their 3D counterparts (in atomic units with $a_0$ denoting the Bohr radius) for both LLH and HHL setups. The radial and transversal trapping frequencies utilized herein are $\omega_r=2\pi \times 65 \, \rm{kHz}$, and $\omega_{\perp}=50 \omega_r$.}
    \label{TAB:scattering}
\end{table}

\section{Summary and Outlook} \label{Sec:Outlook}

The quench dynamics of mass-imbalanced three-body mixtures with either bosonic or fermionic constituents interacting with a third atom is investigated. 
Depending on the mass ratio, we distinguish between the LLH and HHL cases. 
Initially the mixture is confined in a 2D harmonic trap, and assumed to be non-interacting.
The spatial extent of the initial state and the post-quench scattering length are exploited as parameters in order to map out the build-up of two- and three-body correlations via distinct microscopic excitation mechanisms.

In particular, the interactions are abruptly switched on triggering a distinct dynamical response depending on the width of the initial state. A complete knowledge of the energy spectra in conjunction with the fidelity spectrum, allows us to identify the prevalent microscopic mechanisms in terms of specific post-quench eigenstates. 
It is found that if the initial state width is smaller than the three-body harmonic oscillator length $a_{\rm{ho}}$, trimers and atom-dimers contribute predominantly in the dynamics. 
In contrast, for larger widths trap states are those which are significantly populated regardless the mass imbalance of the system.
However, in HHL ensembles for narrow widths, the participation of trimers and atom-dimers prevails in a relatively smaller range of scattering lengths as compared to LLH mixtures.

Interestingly, the participating eigenstates have a distinct imprint on the dynamics of the underlying few-body short-range correlations, as captured by the Tan contacts. 
It is explicated that for an increasing width of the initial state, the magnitude of both the overall  time-averaged two- and the three-body correlations decreases for a fixed 2D scattering length. 
For small widths, these correlations are found to be enhanced as a result of the involvement of trimer states and atom-dimers. 
The respective amplification of the Tan contacts, due to the participation of such states, was also independently reported following the quench dynamics of three-body systems at unitarity in 3D~\cite{colussi_dynamics_2018}. 
Strikingly, for widths larger than the three-body harmonic oscillator length, few-body correlations display sharp peaks at certain scattering lengths. 
This behavior is directly linked to the presence of avoided-crossings among trap and atom-dimer states taking place in the few-body eigenspectrum and signify the non-negligible cooperation of atom-dimers in the time-evolution. 

Overall, our work proposes a scheme to dynamically excite distinct superpositions of eigenstates in three-body mixtures. 
Specifically, it was demonstrated that depending on the interplay between the three-body harmonic oscillator length and the width of the initial state, all three types of eigenstates, that is trimers, atom-dimers and trap states, are possible to be dominantly populated during the nonequilibrium dynamics. 
Moreover, temperature effects are expected  to mitigate few-body correlations as shown in~\cite{yan_harmonically_2013,Bougas_Few_2021}. 
In this sense, the investigation of possible smearing effects of the identified peak structures building upon the time-averaged contacts for large $w>a_{\rm{ho}}$ is a compelling perspective for further research. 

In addition, an interesting question that arises for future studies is how to efficiently populate individual target states, and in particular trimers. 
Their properties such as lifetimes are usually studied indirectly via three-body recombination loss mechanisms~\cite{Pires_Observation_2014,Ulmanis_heteronuclear_2016}. 
However, many questions remain open especially regarding their dynamical formation in a gas~\cite{klauss_observation_2017}.  
A promising route towards achieving this goal would be to utilize time-dependent protocols, in order to activate individual target states instead of superpositions of them generated by quenches. 
There is currently active research for the dynamical creation of the macroscopic population of trimer states in cold gases ~\cite{klauss_observation_2017,Musolino_Bose_2022,colussi_dynamics_2018}.
A first step has already been accomplished in Ref.~\cite{klauss_observation_2017}, where an abrupt tuning of interactions to unitarity and a subsequent sweep to weak repulsion was shown to be able to produce a $8\%$ population of trimers.

\begin{acknowledgments}
G. B. acknowledges financial support by the State Graduate Funding Program Scholarships (Hmb-NFG). 
S.I.M. acknowledges support from the NSF through a grant for ITAMP at Harvard University. 
This work is supported (P.S.) by the Cluster of Excellence `The Hamburg Center for Ultrafast Imaging' of the Deutsche Forschungsgemeinschaft (DFG)-EXC 1074- project ID 194651731. 
The authors thank G. M. Koutentakis for insightful discussions, M.T. Eiles for his comments on the manuscript, and Lydia Schollmeier for the collaboration and discussions in the early stages of this project.
\end{acknowledgments}

\appendix

\section{Adiabatic Hamiltonian and $s$-wave pseudopotential in two-dimensions} \label{Ap:Interactions}

The adiabatic Hamiltonian $H_{{\rm ad}}(R;\boldsymbol{\Omega})$ as introduced in Eq.~\eqref{Eq:hamilt_Hyper} 
is expressed in the following way~\cite{rittenhouse_greens_2010}
\begin{equation}
    H_{{\rm ad}}(R;\boldsymbol{\Omega})=\frac{\hbar^2 \Lambda^2(\boldsymbol{\Omega})}{2\mu R^2}+\frac{3\hbar^2}{8\mu R^2}+\sum_{k} V_{k}(R;\boldsymbol{\Omega}^{(k)}),
    \label{Eq:hamilt_Ad}
\end{equation}
where $\Lambda^2(\boldsymbol{\Omega})$ is the hyperangular operator referring to the centrifugal motion of the three particles~\cite{avery_hyperspherical_1989,Das2016}.
Also, the three-body reduced mass is  $\mu=m_{B/F}/\sqrt{2m_{B/F}/m_X+1}$ with $m_{B/F}$ denoting the mass of the bosons or the fermions depending on the type of the mixture. 

The last term of Eq.~\eqref{Eq:hamilt_Ad} stands for the three (two) pairwise $s$-wave contact interactions among the particles in a BBX (FFX) system. The $V_k$ potential refers to the interaction between the $i$ and $j$ particles (also known as odd-man-out notation where the $i$, $j$ or $k$ indices refer to interaction pairs of the remaining two indices~\cite{rittenhouse_greens_2010}). 
In particular, the $V_k$ interaction is modeled by a 2D pseudopotential which reads~\cite{olshanii_rigorous_2001,kanjilal_coupled-channel_2006}
\begin{eqnarray}
V_{k}(R;\boldsymbol{\Omega}^{(k)})& &=-\frac{\hbar^2 \delta(\alpha^{(k)})}{\mu\sin(2\alpha^{(k)})R^2\ln (A \lambda a^{(k)})} \nonumber \\
& & \times \left[1-\ln \left(A\lambda \sqrt{\mu/\mu_k}R\sin(\alpha^{(k)})\right)\alpha^{(k)}\frac{\partial}{\partial \alpha^{(k)}}\right], \nonumber \\
\label{Eq:Pseudpot}
\end{eqnarray}
where $\alpha^{(k)} \in [0,\pi/2]$ is the hyperangle describing the relative position of two particles compared to the third one. 
For instance, if $\alpha^{(k)}=0$, then the particles $i$ and $j$ are on top of each other, whereas for  $\alpha^{(k)}=\pi/2$, all three particles are collinear. 
Moreover, $\mu_k=\frac{m_im_j}{m_i+m_j}$ is the reduced two-body mass and $A=0.5 \,e^{\gamma}$ with $\gamma\approx0.577$ being the Euler-Mascheroni constant. 
Importantly, $a^{(k)}\equiv a_{ij}$ is the 2D scattering length between the $(i,j)$ pair of particles. The factor $\lambda$ is an ultraviolet-cutoff for the zero-range pseudopotential, setting an upper bound in momentum space. However, it does not affect any observable as argued in Refs.~\cite{olshanii_rigorous_2001,pricoupenko_stability_2007}. 

\section{Hyperangular wave function of the non-interacting initial state} \label{Ap:Hyper_wave}

The hyperangular wave function of the non-interacting initial state (denoted by the $(0)$ superscript) can be expressed~\cite{nielsen_three-body_2001,volosniev_borromean_2014} as follows 
\begin{widetext}
\begin{eqnarray}
\Phi^{(0)}_n(\boldsymbol{\Omega})&=&\sum_{k=1}^3\sum_{\substack{m_1,m_2 \\ \abs{m_1+m_2}=L}}C^{(k)}\mathcal{N}^{(m_1,m_2)}_n\sin^{\abs{m_1}}\alpha^{(k)} \cos^{\abs{m_2}}\alpha^{(k)} Y_{m_1}(\theta_1^{(k)})Y_{m_2}(\theta_2^{(k)}) \frac{\Gamma(1+n+\abs{m_1})}{\Gamma(1+\abs{m_1})n!} \nonumber \\
& & \times _2F_1\left(1+\abs{m_1}+\abs{m_2}+n,-n;\abs{m_1}+1;\sin^2\alpha^{(k)}  \right), \nonumber \\
\label{Eq:Hyperangular_Gaussian}
\end{eqnarray}
\end{widetext}
where $\mathcal{N}^{(m_1,m_2)}_n=\sqrt{\frac{(2n+1+\abs{m_1}+\abs{m_2})\Gamma(n+1)\Gamma(n+1+\abs{m_1}+\abs{m_2})}{2\Gamma(n+1+\abs{m_1})\Gamma(n+1+\abs{m_2})}}$ are normalization coefficients.
The above eigenfunction is the $n$-th eigenstate ($n$ is a non-negative integer) of the hyperangular operator $\Lambda^2(\boldsymbol{\Omega})$~\cite{avery_hyperspherical_1989,Das2016} with eigenvalues $\lambda_n(\lambda_n+2)$  where 
\begin{equation}
    \lambda_n=2n+\abs{m_1}+\abs{m_2},
    \label{Eq:Momentum_eigenvalues}
\end{equation}
and $L=\abs{m_1+m_2}$ being the total angular momentum of the three-body system.
It is expressed in terms of the angular quantum numbers $m_1$, $m_2$ related to the polar angles $\theta_1^{(k)}$ and $\theta_2^{(k)}$. 
The polar angles $\theta_1^{(k)}$ and $\theta_2^{(k)}$ refer to the orientation of the Jacobi vectors $\boldsymbol{\rho}_1^{(k)}$, $\boldsymbol{\rho}_2^{(k)}$ in the 2D plane, respectively, where $\boldsymbol{\rho}_1^{(k)}$ is the relative distance of the $(i,j)$-pair and $\boldsymbol{\rho}_2^{(k)}$ is the relative vector of the $k$ spectator particle relative to the $(i,j)$-pair's center of mass. 
The summation running over these angular quantum numbers is restricted by the condition $L=\abs{m_1+m_2}$. Note that in the case of three identical particles, $n=1$ gives an unphysical solution and therefore it is not allowed~\cite{Incao_adiabatic_2014}.
Additionally, $_2F_1(a,b;c;\cdot)$ is the Gauss hypergeometric function~\cite{abramowitz_handbook_1965} and $Y_m(x)=e^{i m x}/\sqrt{2\pi}$ are plane waves. 
The angle $\alpha^{(k)}$ determines the ratio of the measure of the two Jacobi vectors via the relation $\tan \alpha^{(k)}=\rho_1^{(k)}/\rho_2^{(k)}$ [see also Appendix~\ref{Ap:Interactions}]. 

The particle statistics of the above wave function is properly taken into account by the first summation and the $C^{(k)}$ coefficients. These read explicitly $(C_1,-C_1,0)$ and $(C_1,C_1,C_2)$ for FFX and BBX systems respectively, with the $C_1$ and $C_2$ terms being normalization coefficients. 
The hyperangular wave functions $\Phi_{\nu}(R;\boldsymbol{\Omega})$, (which are eigenstates of $H_{\rm{ad}}(R;\boldsymbol{\Omega})$) correspond to the interacting post-quench eigenstates and have angular quantum numbers $(m_1,m_2)=(0,\pm L)$ due to the $s$-wave zero-range pseudopotential. 
As such, the relevant subset in the summation [Eq.~\eqref{Eq:Hyperangular_Gaussian}] will also be $(0, \pm L)$. 
Indeed, the remaining terms in the summation have a zero contribution in the overlap coefficients, $c_{\rm{f},\rm{in}}$, since the plane-waves $Y_m(\cdot)$ are orthonormal. 
Here, we focus on $n=0$, that is the ground state. Note that the hyperangular wave function does not depend on the hyperradius $R$ since in the non-interacting case $H_{\rm{ad}}(R;\boldsymbol{\Omega})$ does not depend on $R$, as all interaction terms $V_k(R;\boldsymbol{\Omega}^{(k)})$ drop [see also Appendix \ref{Ap:Interactions}].   

\section{Quench dynamics of the LLH BBX mixture for initial states with $w=a_{\rm{ho}}$}  \label{Ap:Same_width}

\begin{figure}[t!]
\centering
\includegraphics[width=0.4\textwidth]{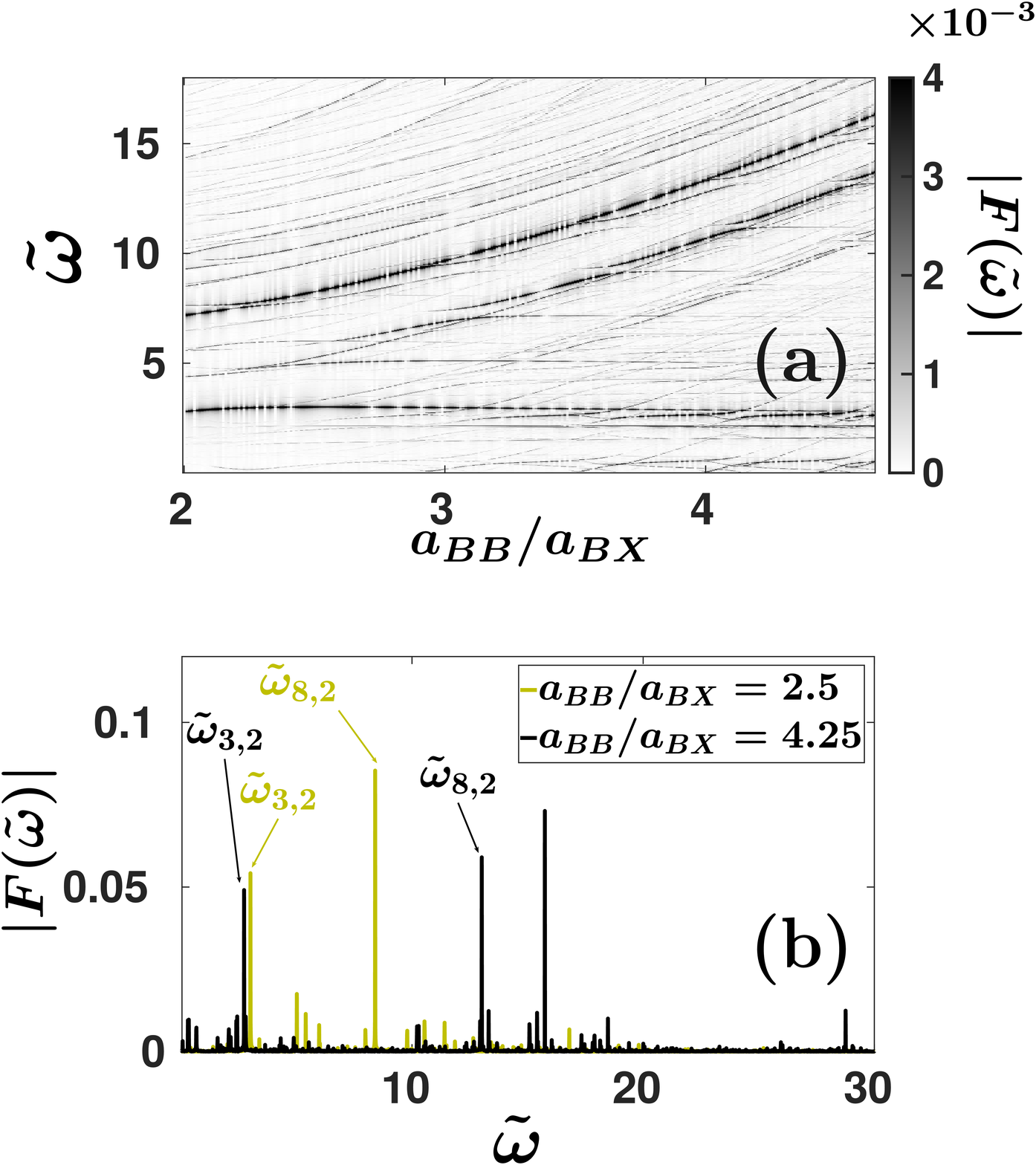}
\caption{(a) Fidelity spectrum ($\abs{F(\tilde{\omega})}$) for the LLH BBX system subjected to a quench of $a_{BB}/a_{BX}$ from an initial non-interacting state with $w/a_{\rm{ho}}=1$.  
(b) Profiles of $\abs{F(\tilde{\omega})}$ at different scattering length ratios $a_{BB}/a_{BX}$ (see legend).  
The excitation processes involve majorly trimer and atom-dimer states which are imprinted in the spectrum as branches that are sensitive to the scattering length. 
Notice that the participation of the second trimer is reduced compared to the $w/a_{\rm{ho}}=0.78$ case, resulting in different branches than in $\abs{F(\tilde{\omega})}$ depicted in \cref{Fig:Fid_Spec_LLH} (c).} 
\label{Fig:Spec_Fid_LLH_Ap}
\end{figure}

For completeness, we shall also analyze the excitation spectrum of three-body mixtures starting from a pre-quench state of width $w=a_{\rm {ho}}$. 
As characteristic system for this investigation we consider a LLH BBX system whose fidelity spectrum [Eq.~\eqref{Eq:Fid_Spec}] is illustrated in \cref{Fig:Spec_Fid_LLH_Ap} for varying post-quench  $a_{BB}/a_{BX}$. 

Recall that for $w/a_{\rm{ho}}=0.78$, the second trimer state $\rm{f}=2$ contributes the most in the quench dynamics of the LLH BBX setting, see also the discussion in Sec.~\ref{Sec:LLH_small_width}. 
The predominant population of the second trimer yields, in particular, excitation branches that are strongly influenced by $a_{BB}/a_{BX}$ [\cref{Fig:Fid_Spec_LLH} (a)]. 
This is a consequence of the fact that the branches associated to these transitions refer to energy differences between the $\rm{f}=2$ trimer and the trap states and are increasing as $a_{BB}/a_{BX}$ is tuned to larger values. 

These excitation branches are still present even for an initial state width $w/a_{\rm{ho}}=1$ as shown in \cref{Fig:Spec_Fid_LLH_Ap} (a).   
Here, the almost constant frequency branch located around $\tilde{\omega} \simeq 2$, stemming from the transition among the second trimer ($\rm{f}=2$) and the first atom-dimer ($\rm{f}=3$) states, is more enhanced than in the case where $w/a_{\rm{ho}}=0.78$ (compare $\tilde{\omega}_{3,2}$ in \cref{Fig:Spec_Fid_LLH_Ap} (b) and \cref{Fig:Fid_Spec_LLH} (c)). 
This difference is attributed to the fact that the occupation of the first atom-dimer state is larger when $w=a_{\rm{ho}}$, while the one from the second trimer is reduced, a result that is supported by the corresponding overlap coefficients $c_{\rm{f},\rm{in}}$. 
To be more precise, the population of the $\rm{f}=2$ trimer as long as $w/a_{\rm{ho}}=0.78$ ($w/a_{\rm{ho}}=1$) ranges from $73 \%$ ($57\%$) to $35 \%$ ($21\%$) within the interval $a_{BB}/a_{BX} \in [2,4.6]$. 
Apart from the enhanced population of the first atom-dimer, the contribution of trap states, similar to the ones populated also for $w/a_{\rm{ho}}=0.78$, increases as well with respect to $w/a_{\rm{ho}}=0.78$. 
This is imprinted in the spectrum by the larger number of faint excitation branches, compare in particular \cref{Fig:Spec_Fid_LLH_Ap} (a) where $w/a_{\rm{ho}}=1$ with \cref{Fig:Fid_Spec_LLH} (a) for which $w/a_{\rm{ho}}=0.78$. 

Similar observations to the above can be made for the other type of mixtures utilized in the main text. 
Regarding the LLH FFX system, the contribution of the first two atom-dimer states at $w/a_{\rm{ho}}=1$ remains the same in comparison to $w/a_{\rm{ho}}=0.78$ for $1/a_{FX}<1$. 
Otherwise, it reduces further from the value obtained for $w/a_{\rm{ho}}=0.78$ ($18\%$ versus $25 \%$ at $1/a_{FX}=4.5$).
This reduction is compensated by an increasing population of a few trap states. 
Due to the reduced number of participating post-quench eigenstates compared to smaller $1/a_{FX}$, the time-averaged fidelity possesses a smaller magnitude for $1/a_{FX}>3$ [see \cref{Fig:Fid_Av_LLH} (b) for $w/a_{\rm{ho}}=1$]. 
In a similar way, the population of trimers and first atom-dimers also drops when considering $w/a_{\rm{ho}}=1$ for the HHL mixtures (both BBX and FFX systems) as compared to the scenario where $w/a_{\rm{ho}}=0.57$. 

\bibliography{Three_Bodies_dynamics.bib}

\end{document}